\documentclass{ws-procs961x669} 
\begin{document}
\def\be{\begin{equation}}
\def\ee{\end{equation}}
\def\bea{\begin{eqnarray}}
\def\eea{\end{eqnarray}}
\def\fr{\frac}
\def\p{\partial}
\def\a{\alpha}
\def\b{\beta}
\def\g{\gamma}
\def\d{\delta}
\def\e{\eta}
\def\r{\rho}
\def\th{\theta}
\def\ph{\phi}
\def\eps{\epsilon}
\def\fr{\frac}
\def\l{\label}
\def\o{\omega}
\def\om{\omega}
\def\O{\Omega}
\def\epsilon{\varepsilon}
\newcommand{\dd}{\mbox{d}}
\newcommand{\veps}{\varepsilon}

\title{The world of long-range interactions: A bird's eye view}
\author{Shamik Gupta}
\address{Department of Physics, Ramakrishna Mission Vivekananda
University, \\ Belur Math, Howrah 711 202, West Bengal, India \\
E-mail: shamikg1@gmail.com}

\author{Stefano Ruffo}
\address{
SISSA, INFN and ISC-CNR,\\
Via Bonomea 265, I-34136 Trieste, Italy \\
E-mail: ruffo@sissa.it}
\begin{abstract}
In recent years, studies of long-range interacting (LRI) systems have taken
centre stage in the arena of statistical mechanics and dynamical
system studies, due to new theoretical developments involving tools from as diverse
a field as kinetic theory, non-equilibrium statistical mechanics, and large deviation
theory, but also due to new and exciting experimental
realizations of LRI systems. In the first, introductory, Section 1, we
discuss the general
features of long-range interactions, emphasizing in particular the main 
physical phenomenon of non-additivity, which leads to a plethora of distinct
effects, both thermodynamic and dynamic, that are not observed with short-range interactions: Ensemble inequivalence, slow relaxation, broken ergodicity. In Section 2, we discuss several physical systems with long-range interactions: mean-field spin systems, self-gravitating systems, Euler equations in
two dimensions, Coulomb systems, one-component electron plasma, dipolar systems,
free-electron lasers. In Section 3, we discuss the general scenario of
dynamical evolution of generic LRI systems. In Section 4, we discuss an
illustrative example of LRI systems, the Kardar-Nagel spin system, which involves discrete degrees of
freedom, while in Section 5, we discuss a paradigmatic example involving
continuous degrees of freedom, the so-called Hamiltonian mean-field
(HMF) model. For the former, we demonstrate the
effects of ensemble inequivalence and slow relaxation, while for the HMF
model, we emphasize in particular the occurrence of
the so-called quasistationary states (QSSs) during relaxation towards
the Boltzmann-Gibbs equilibrium state. The QSSs are non-equilibrium
states with lifetimes that diverge with the system size, so that in the
thermodynamic limit, the systems remain trapped in the QSSs, thereby
making the latter the effective stationary states. In Section 5, we also discuss an experimental
system involving atoms trapped in optical cavities, which may be
modelled by the HMF system. In Section 6, we address the issue of ubiquity of the quasistationary behavior by
considering a variety of models and dynamics, discussing in each case
the conditions to observe QSSs. In Section 7, we investigate the
issue of what happens when a long-range system is driven out of
thermal equilibrium. Conclusions are drawn in Section 8.
\end{abstract}

\keywords{Long-range interactions; Non-additivity; Ensemble
inequivalence; Slow relaxation; Quasi-stationary states.}
\bodymatter

\section{Introduction: General considerations}
\l{sec:introduction}

In this Section, we discuss the generalities of long-range interacting
systems. A detailed discussion, with extensive lists of references,
may be found in several recent articles and books, see Refs. \cite{Dauxois:2002,Campa:2008,Ruffo:2008,Campa:2009,Dauxois:2009,Dauxois:2010,Bouchet:2010,Campa:2014}. More
recent works discussed in the later parts of this article are covered in Refs.
\cite{Gupta:2010-1,Gupta:2010-2,Gupta:2013,Gupta:2011,Barre:2014,Schutz:2014,Schutz:2015,Jager:2016,Casetti:2014,Teles:2015,Teles:2016,Gupta:2016}.

Long-range interacting (LRI) systems are those in which the two-body
interparticle potential decays at large separation $r$ as
\be
V(r) \sim \frac{J}{r^\alpha};~0 \leq \alpha \leq d,
\l{eq:Vr}
\ee
where $d$ is the dimension of the embedding space, and $J$ is the
coupling strength \footnote{
An alternative classification, based on dynamical considerations,
namely, the conditions for the existence of the so-called
quasistationary states in LRI systems, is proposed in A. Gabrielli, M.
Joyce, and J. Morand, Phys. Rev. E {\bf 90}, 062910 (2014).}. The range of allowed values of the decay exponent
$\alpha$ implies that the energy per particle, $\varepsilon$, scales
super-linearly with the system size. This feature is easily demonstrated by
considering the example of a particle placed at the center of a
hypersphere of radius $R$ in $d$ dimensions, with the other particles homogeneously distributed
with a mass density $\rho$. For such a system, considering the
interaction potential (\ref{eq:Vr}), the energy per particle is given as 
\be
\varepsilon=\int_{\delta}^R {\rm d}^d r ~\rho \frac{J}{r^\alpha}=\frac{\rho J
\Omega_d}{d-\alpha}\left[R^{d-\alpha}-\delta^{d-\alpha}\right],
\l{eq:epsilon}
\ee
where $\delta \to 0$ is a short distance cut-off introduced to exclude the
contribution to the energy due to particles located in a small
neighborhood of radius $\delta$, and is motivated by the need to
regularize the divergence of the potential (\ref{eq:Vr}) at short 
distances. In Eq. (\ref{eq:epsilon}), $\Omega_d$ denotes the angular volume in $d$
dimensions. From the equation, it follows that as $R$ is increased, the
energy $\varepsilon$ remains finite for $\alpha > d$, implying thereby the
linear scaling of the total energy $E$ with the volume $V\sim R^d$,
thus making the system extensive. These systems are called short-range
interacting systems. On the other hand, for our allowed values of
$\alpha$, the energy $\varepsilon$ scales with the volume as $\varepsilon
\sim V^{1-\alpha/d}$ (the energy scales logarithmically with $V$ in the marginal case $\alpha =
d$), thereby implying a super-linear scaling of the total energy of LRI
systems with 
$V$, as $E \propto V^{2-\alpha/d}$. The LRI systems are thus generically
non-extensive. For such systems, on computing the free energy $F \equiv E-TS$, with
$T$ being the intensive temperature and $S$ being the entropy that typically scales linearly with the volume, $S
\sim V$, we find due to the super-linear scaling of $E$ with $V$ that the thermodynamic properties are dominated by the
energy. In particular, the equilibrium state of a mechanically
isolated LRI system at constant temperature, obtained by minimizing $F$, corresponds to the one
with the minimum energy, that is, the ground state, allowing for no
thermal fluctuations. Of course, in reality, there ought to be a
competition between the energy and the entropy contributions to the free
energy in order to
have such phenomena as phase transitions that are known to occur in
LRI systems. A way out from this energy dominance consists in scaling
the coupling constant as
\be
J \to \frac{J}{V^{1-\alpha/d}},
\l{eq:J-scaling}
\ee
thereby making the energy extensive in the volume. Note that this is
just a ``mathematical trick" (Kac's trick) to properly study LRI systems within the
framework of equilibrium statistical mechanics that exists for
short-range ones, and does not correspond to any physical effect.
Indeed, no interaction whose strength changes on varying the
volume is known to occur. Applying this trick, one can obtain the free energy
per particle, and then revert to the actual physical description by scaling back the
coupling constant. An equivalent alternative  to Kac's trick, which still allows
for an effective competition between the energy and the entropy
contribution to the free energy, consists in
rescaling the temperature as
\be
T \to \frac{T}{V^{\alpha/d-1}}.
\l{eq:T-scaling}
\ee

Beyond the rescaling procedures discussed above, which were implemented to obtain
a meaningful large-volume limit for LRI systems and competing energy and
entropy contributions to the free energy, let us illustrate how such a
competition may actually occur in nature, by
considering a relevant LRI system in the arena of astrophysics, namely, that of
globular clusters, see Fig. \ref{fig:globular}. These clusters are
gravitationally bound concentrations of $N \sim 10^4-10^6$ 
stars that are spread over a volume that has a diameter ranging from several tens to about
$200$ light years (1 light year = $9.4 \times 10^{15}$ m). For a
typical globular cluster (M2), one has $N=1.5 \times 10^5$, $R=175$
light years, and total mass $M=2 \times 10^{30}$ Kg. An order-of-magnitude estimate of
energy and entropy may be done as follows:
\be
E=\frac{GN^2M^2}{R},~S=k_B N \implies \frac{E}{S} \sim
\frac{GNM^2}{k_BR} \sim 1.7 \times 10^{60}~{\rm K},
\l{eq:globular-clusters}
\ee
where $G$ is the gravitational constant, and $k_B$ is the Boltzmann constant.
To such an extremely high temperature as $\sim 10^{60}$ K, one can
associate a velocity by invoking energy equipartition
(neglecting interactions), as $v=\sqrt{\frac{3k_BT}{M}} \simeq 5.9$
Km/s. Typical star velocities indeed range between a few Km/s  to about
$100$ Km/s. Thus, for systems such as these for which the temperature is high
enough, the energy, although super-linear
in volume ($E \sim V^{5/3}$), can effectively compete with the entropy contribution to the
free energy.

\begin{figure}[ht!]
\centering
\includegraphics[width=50mm]{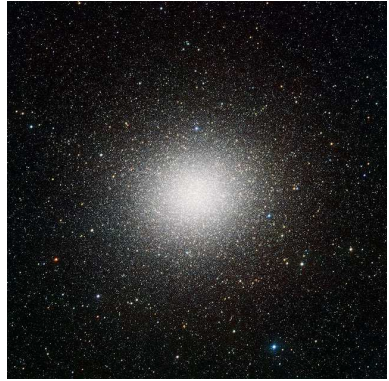}
\caption{Spherically symmetric mass distribution of stars in a globular cluster.}
\l{fig:globular}
\end{figure}

Let us now make an important remark: Although Kac's trick allows to
obtain an energy that is extensive in the volume, it is not necessarily
additive (additivity implies extensivity, but not the converse). A
simple example will illustrate the point. Consider the well-studied
Curie-Weiss model of magnetism, with the Hamiltonian given by
\be
H_{\rm CW}=-\frac{J}{2N} \sum_{1 \le i < j \le N} \sigma_i \sigma_j,
\l{eq:H-CW}
\ee
where $\sigma_i=\pm 1$ are spin variables occupying the sites $i$ of a
lattice. The model mimics a mean-field
system (every spin interacting with every other with the same strength),
which may be considered as the $\alpha \to 0$ limit of the potential
(\ref{eq:Vr}). 
Being a mean-field system, one does not need to specify the structure of
the underlying lattice, excepting to mention that every site is connected to every
other. In the Hamiltonian (\ref{eq:H-CW}), the coupling strength has been rescaled by using Kac's
trick, so that the energy is extensive in the number of spins given by $N$. 
Let us consider a macrostate with zero total magnetization:
$M \equiv \sum_{i=1}^N\sigma_i=0$, which is composed of $N/2$ spin-$(+1)$ sites
and $N/2$ spin-$(-1)$ sites, see Fig. \ref{fig:additive}. Since the energy is
proportional to the square of magnetization (see Eq. (\ref{eq:H-CW})),
the total energy of the system is $E_{1+2}=0$. However,
the energy of the two parts, namely, $E_1=E_2=-J/8N$, does not
vanish, and, therefore, $E_{1+2} \neq E_1+E_2$.

\begin{figure}[ht!]
\centering
\includegraphics[width=70mm]{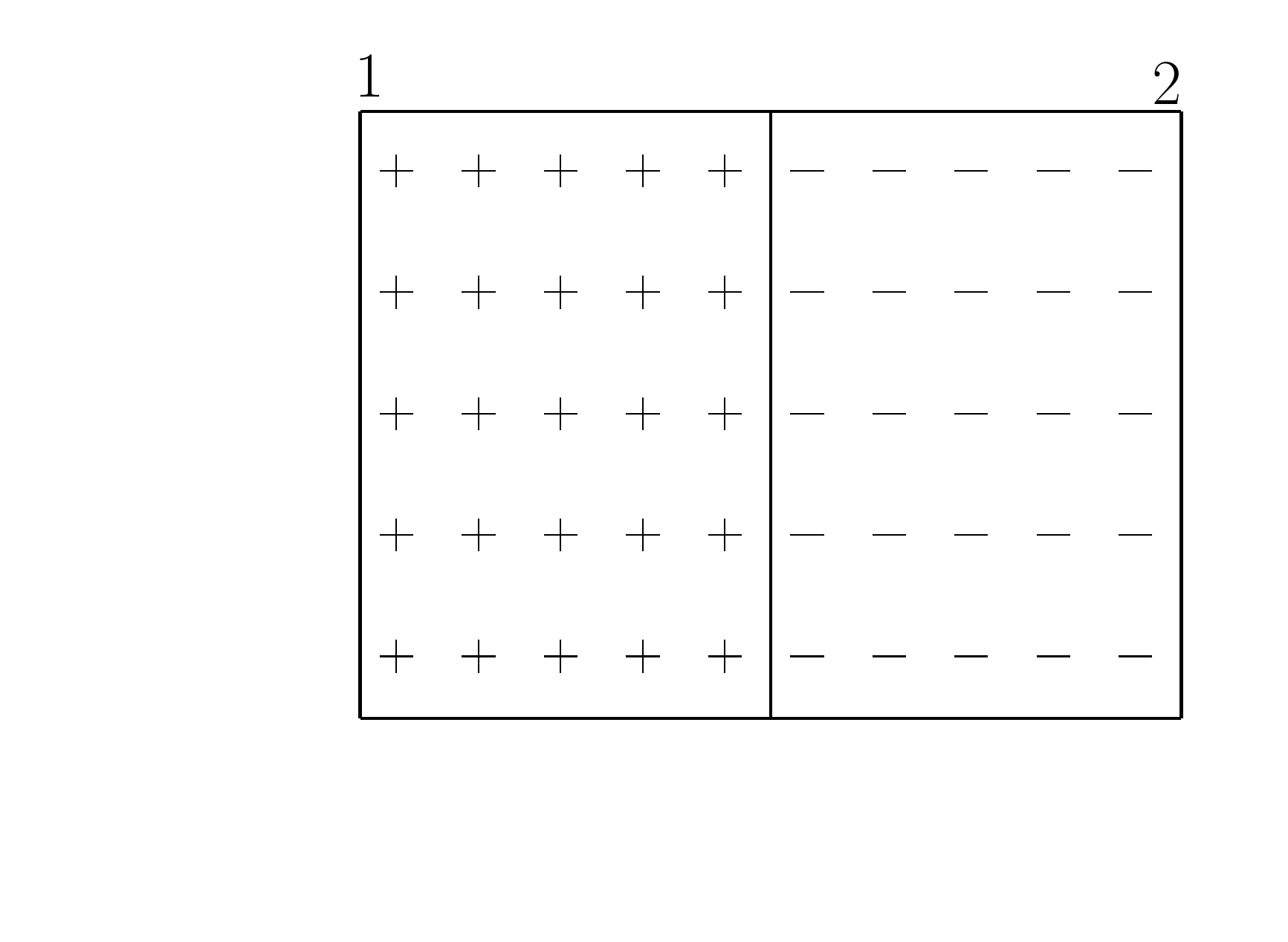}
\caption{Non-additivity in the extensive Curie-Weiss model: A zero magnetization macrostate of the Curie-Weiss system, Eq.
(\ref{eq:H-CW}), constituted by $N/2$ up-spins
in domain 1 and $N/2$ down-spins in domain 2. Here, an up-spin is
denoted by a $+$ sign and a down-spin by a $-$ spin.}
\l{fig:additive}
\end{figure}

As it will emerge in the rest of this article, the violation of
additivity will be crucial in determining both thermodynamic and dynamic
properties of LRI systems, making them quite distinct from short-range
ones. This point will be demonstrated by considering several
illustrative examples in the later parts of the paper. As a warm-up, we
may mention that a violation of additivity implies a violation of
convexity of the domain of accessible macrostates of an LRI system, for
example, a magnetic one in the magnetization ($M$) - energy ($E$) plane.
An example of such a violation is shown in Fig. \ref{fig:convexity}, where the boundary of the region of
accessible macrostates is shown to have the shape of a bean. For short-range systems for which additivity is satisfied,
standard thermodynamics implies that all states satisfying
\be
E=\lambda E_1+(1-\lambda) E_2,~M=\lambda M_1+(1-\lambda) M_2,~0 \leq \lambda \leq 1
\l{eq:convexity}
\ee
must occur at the macroscopic level; this is in general not the case for
LRI systems, and may imply a violation of ergodicity in the microcanonical 
ensemble. For example, the states $(M_1,E_1)$ and $(M_2,E_2)$ in Fig.
\ref{fig:convexity} are not connected by any continuous energy-conserving dynamics.

\begin{figure}[ht!]
\centering
\includegraphics[width=60mm]{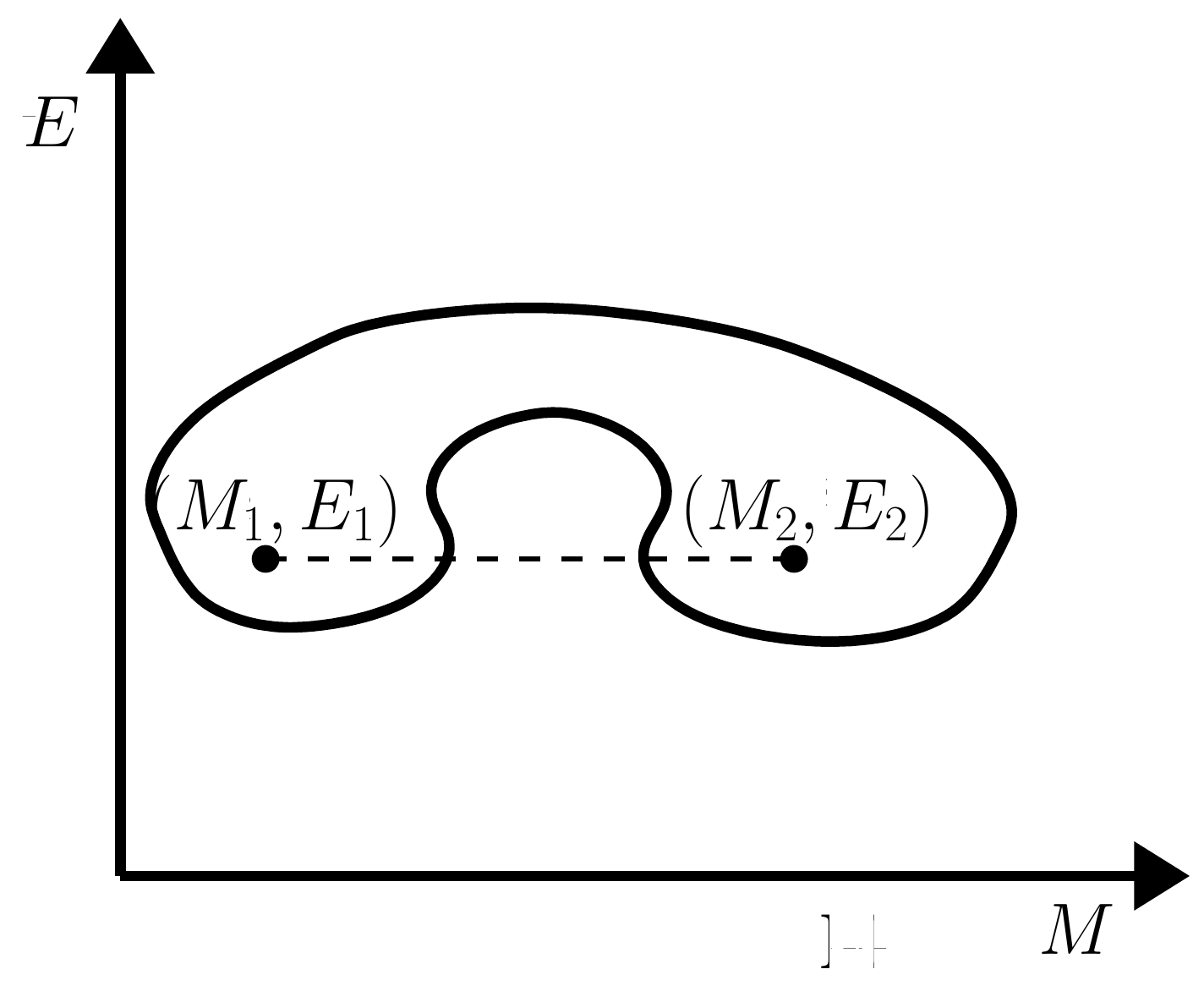}
\caption{Non-convex shape of the region of accessible macrostates in the
magnetization-energy plane for a magnetic LRI system. The states
$(M_1,E_1)$ and $(M_2,E_2)$ are not connected by any continuous
energy-conserving dynamics.}
\l{fig:convexity}
\end{figure}

On account of the violation of additivity, one should exercise caution in
discussing equilibrium properties of LRI systems using the canonical
ensemble whose derivation from the microcanonical ensemble relies on
holding of additivity. Let us briefly recall the derivation. 
The microcanonical partition function for a system of $N$ particles
contained in a volume $V$ in
$d=3$ dimensions is given by 
\be
\Omega (E,V,N) \propto \int {\rm d}^{3N} q {\rm d}^{3N} p~\delta(E-H(p,q)),
\l{eq:microcanonical}
\ee
where $(q,p)$ are the canonically conjugate variables, and $H$ is the
Hamiltonian. The entropy is defined via 
\be
S(E,V,N) = k_B \ln \Omega(E,V,N),
\l{eq:entr}
\ee
where an energy scale should be included in the logarithm to make its
argument dimensionless. In deriving the canonical ensemble for a
short-range system, one considers an isolated macroscopic system with
energy $E$ that is composed of a ``small" part (the subsystem
of interest) with energy equal to $E_1$, volume equal to $V_1$ and
number of particles equal to $N_1$, and a ``large" part that plays the role
of a ''bath", with energy equal to $E_2 \gg E_1$, volume equal to $V-V_1 \gg V_1$ and number of
particles equal to $N-N_1 \gg N_1$. The additivity of energy implies that one has
$E_2=E-E_1$, so that the
probability distribution $p(E_1)$ that the ``small'' system
has energy $E_1$ is given by
\be
p(E_1) \propto \Omega_2(E-E_1,V-V_1,N-N_1).
\l{eq:canonical-ensemble-derivation-0}
\ee
Using the definition of entropy and a Taylor expansion, one gets
\bea
p(E_1)&=&\exp \left[S_2(E-E_1)\right] \nonumber
\\
& \approx & \exp \left[S_2(E)-E_1\left.\frac{\partial
S_2}{\partial E}\right|_E + \cdots \right] \nonumber
\\
& \propto &\Omega_2(E,V-V_1,N-N_1)\;e^{-\beta E_1},
\l{eq:canonical-ensemble-derivation}
\eea
where $S_2(E) \equiv S_2(E,V-V_1,N-N_1)$, and $\beta \equiv \left.\frac{\partial S_2}{\partial E}\right|_E$ 
is the inverse temperature: $\beta=1/(k_BT)$. Equation
(\ref{eq:canonical-ensemble-derivation}) is the usual canonical ensemble
description for the energy distribution of the system of interest. In
describing LRI systems, which we have shown to be generically
non-additive so that the derivation leading to
(\ref{eq:canonical-ensemble-derivation}) does not hold, we will consider both the microcanonical
description, Eq. (\ref{eq:microcanonical}), and the canonical one, Eq.
(\ref{eq:canonical-ensemble-derivation}). For the latter, however, we will adopt an alternative
physical interpretation, namely, that of the system of interest in
interaction with an external heat bath at temperature $T$ that induces stochastic fluctuations
into the dynamics of the system. 

Thermodynamic ensembles could be inequivalent for LRI systems \cite{Kiessling:1997,Barre:2001,Ellis:2002,Pikovsky:2014}: a macroscopic physical state that is realizable
in one ensemble is not realized in the other. We now discuss this point.
Ensemble equivalence in the thermodynamic limit is mathematically based on certain properties
of the partition functions. The thermodynamic limit corresponds to considering simultaneously the limits $N\rightarrow \infty$,
$E\rightarrow \infty$ and $V\rightarrow \infty$, such that one has $N/V
\rightarrow n$ and $E/N \rightarrow \varepsilon$, where the particle density
$n\ge 0$ and the energy per particle $\varepsilon$ are finite
quantities. In this limit, the entropy per particle is given by 
\be
s(\varepsilon,n) \equiv \lim_{N\rightarrow \infty} \frac{1}{N}S(E,V,N).
\l{eq:entrn}
\ee
The function $s(\varepsilon,n)$ is continuous, increasing with $\varepsilon$ at a
fixed $n$, so that the temperature $T=\left(\partial s/\partial
\varepsilon\right)^{-1}$ is a positive quantity. For short-range systems, $s(\varepsilon,n)$
turns out to be a concave function of $\varepsilon$ at a fixed
$n$:
\be
s\left(\lambda\varepsilon_1 + (1-\lambda)\varepsilon_2,n\right)
\ge \lambda s(\varepsilon_1,n) +(1-\lambda) s(\varepsilon_2,n),
\l{eq:concpro}
\ee
for any choice of $\varepsilon_1$ and $\varepsilon_2$, with $0\le
\lambda \le 1$. The partition function in the canonical ensemble is given by
\be
Z(\beta,V,N) \equiv \int {\rm d} q^{3N} {\rm
d}p^{3N}~\exp \left[ -\beta H(p,q)\right].
\l{eq:canon}
\ee
In the thermodynamic limit, the free energy per particle is obtained as
\be
f(\beta,n)\equiv-\frac{1}{\beta}\lim_{N\rightarrow \infty} \frac{1}{N}
\ln Z(\beta,V,N).
\l{eq:freen}
\ee
Moreover, at a fixed $n$, the function $\phi(\beta,n)\equiv \beta
f(\beta,n)$ (the rescaled free energy) is concave in $\beta$. 
The equivalence between the microcanonical and the canonical ensemble is a
consequence of the concavity of $\phi$ and $s$ and of the relation
between these two functions given by the Legendre-Fenchel Transform
(LFT). Indeed, one can easily prove that $\phi(\beta,n)$ is the LFT of
$s(\varepsilon,n)$:
\be
\phi (\beta,n)=\inf_{\varepsilon}
\left[\beta \varepsilon - s(\varepsilon,n)\right],
\l{eq:legen1}
\ee
and also the inverse LFT holds, since $s(\varepsilon,n)$ is concave in
$\varepsilon$: 
\be
s(\varepsilon,n)= \inf_{\beta} \left[\beta
\varepsilon - \phi(\beta,n)\right].
\l{eq:legen2}
\ee
These relations prove ensemble equivalence, because for each value of
$\beta$, there is a value of $\varepsilon$ that satisfies
Eq. (\ref{eq:legen1}), and, conversely, for each value of $\varepsilon$, 
there is a value of $\beta$ satisfying Eq. (\ref{eq:legen2}).
Figure \ref{fig:lft} provides a visual explanation of the
relation between $s$ and $\phi$ and of the correspondence between
$\varepsilon$ and $\beta$. Note that at a first-order phase transition,
the entropy has a constant slope in the energy range
$[\varepsilon_1,\varepsilon_2]$ (the phase coexistence region), resulting in a free 
energy with a cusp at the transition inverse temperature $\beta_t$, see
Fig. \ref{fig:firstorder}.

\begin{figure}[ht!]
\centering
\includegraphics[width=90mm]{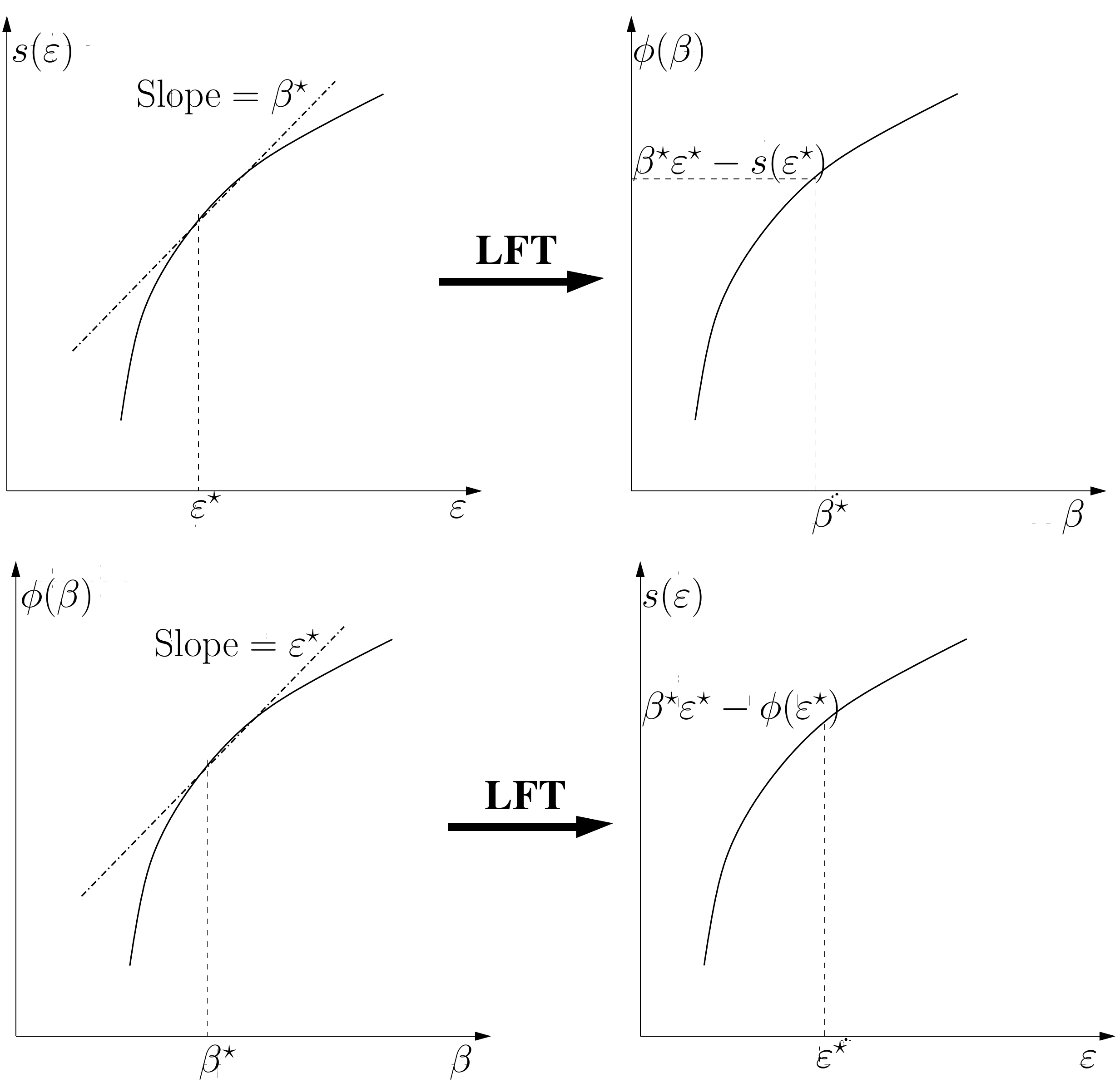}
\caption{(Upper panel) Free energy from entropy by performing a
Legendre-Fenchel transform. (Lower
panel) Entropy from free energy by the same transform.}
\l{fig:lft}
\end{figure}

\begin{figure}[ht!]
\centering
\includegraphics[width=90mm]{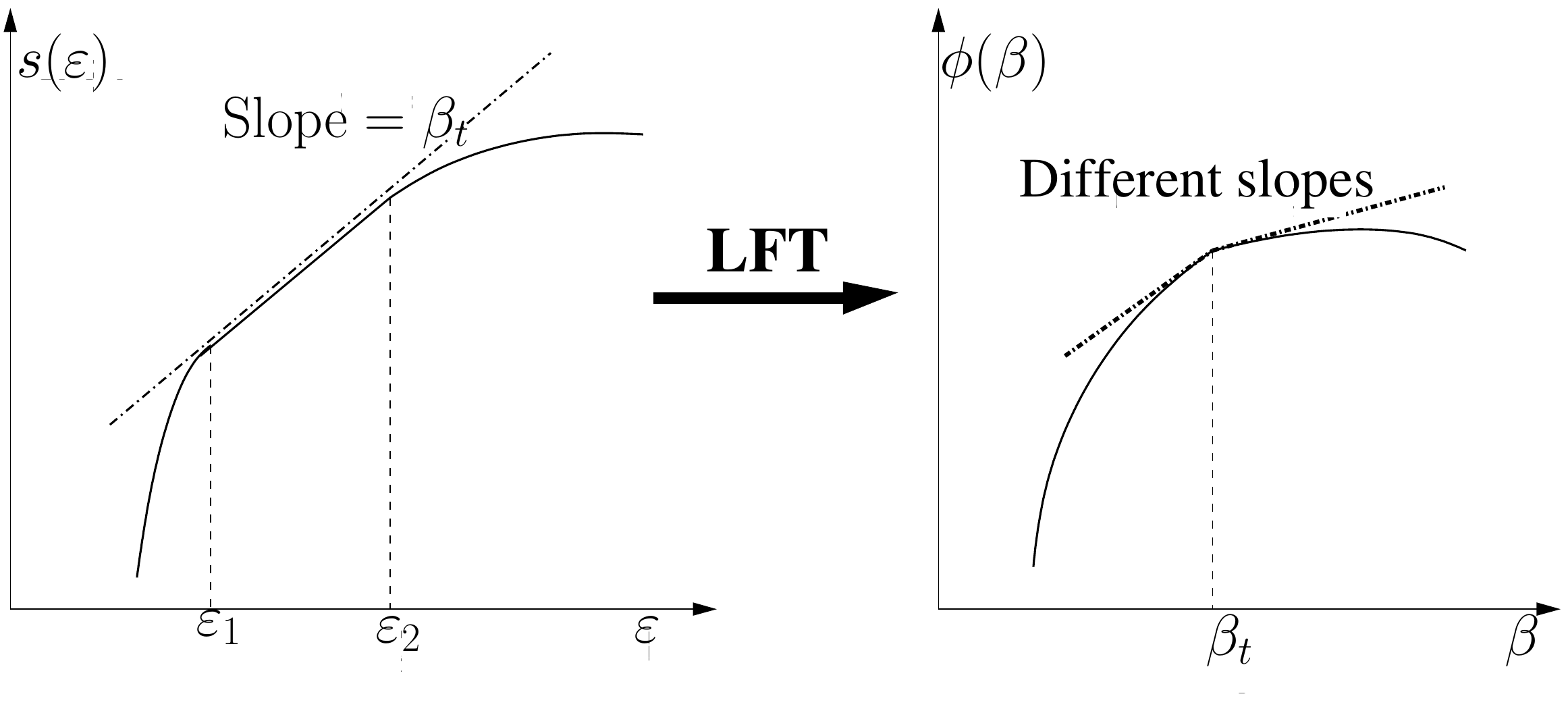}
\caption{Relation between the entropy and the free energy at a first-order phase transition.}
\l{fig:firstorder}
\end{figure}

Now, for LRI systems, the entropy may be a non-concave function of the
energy. In this case, the Legendre-Fenchel transform is no more
involutive: if applied to the entropy, it returns the correct free energy. However, the Legendre-Fenchel
transform of the free energy does not coincide with the entropy, but rather with its
concave envelope; this is the basic feature causing ensemble inequivalence.

\section{Examples of LRI systems}
\l{sec:examples}

A wide class of LRI systems comprises $N$ interacting particles having the total potential energy
\be
U(\vec{r}_1,\dots,\vec{r}_N)=
\sum_{1\le i<j\le N} V(|\vec{r}_i -\vec{r}_j|)+ 
\sum_{i=1}^N V_e(\vec{r}_i),
\l{eq:U}
\ee
where $\vec{r}_i$ is the position of the $i$-th particle, $V$ is the interparticle potential
and $V_e$ represents the potential energy due to an external field. In contrast
to the continuum description of Eq. (\ref{eq:U}), long-range
interactions may also be defined on a lattice (the Curie-Weiss model
considered above was one such system), with the potential energy having the
form
\be
U({\bf q}_1,\dots,{\bf q}_N)=\sum_{1\le i < j \le
N}C_{ij}V({\bf q}_i,{\bf q}_j) + \sum_{i=1}^N V_e({\bf q}_i),
\l{eq:potlat}
\ee
where ${\bf q}_i$ represents the ``internal" degrees of freedom
occupying the lattice site $\mathbf{r}_i$, and the coupling given by 
\be
C_{ij}=\frac{1}{|\vec{r}_i -\vec{r}_j|^\alpha};~ 0 \leq \alpha \leq d 
\l{eq:Cij}
\ee
bears the long-range nature of the interaction between the particles.

A model with the Hamiltonian of the type (\ref{eq:potlat}) is the
so-called Dyson model, comprising Ising spins $\sigma_i=\pm1$ occupying
the sites of a one-dimensional lattice with $N$ sites. The Hamiltonian
is given by  
\be
H_{\rm Dyson}= -\frac{J}{2}  \sum_{1 \le i < j \le N} \frac{\sigma_i
\sigma_j}{|i - j|^{1+\sigma}}.
\l{eq:Dyson}
\ee
The scaling properties of the energy are $E \sim N$ for $\sigma > 0$,
and $E \sim N^{1 - \sigma}$ for $-1 \leq \sigma \leq 0$. The model
exhibits a ferromagnetic phase transition for $0 < \sigma \leq 1$, and
no phase transition for $\sigma >1$. At $\sigma=1$, a jump in the magnetization at the transition
point together with a diverging correlation length (which are signatures
of the so-called
mixed-order phase transitions) occur. For $-1 \leq \sigma \leq 0$, in
accordance with our discussions in the preceding Section, one can
apply Kac's trick, $J \to J N^{\sigma}$, to obtain a free
energy that is extensive in $N$. 

An example of the type (\ref{eq:U}) is afforded by the most notable and
fundamental system of long-range interaction, namely, that of a
self-gravitating system, for which the potential energy is given by
\be
U(\vec{r}_1,\dots,\vec{r}_N)=-Gm^2
\sum_{1 \le i<j\le N} \frac{1}{|\vec{r}_i
-\vec{r}_j|}.
\l{eq:U-gravity}
\ee
In order to get a microcanonical partition sum that is finite, one needs
to confine the system to a box of finite volume $V$, as is also the case
for doing statistical mechanical calculations of the ideal gas. We thus have
\be
\Omega(E,N,V) = \int_V \prod_{i=1}^N {\rm d} \vec{r}_i {\rm d} \vec{p}_i~  
\delta (E-K-U) \propto \int_V \prod_{i=1}^N {\rm d} \vec{r}_i~ (E-U)^{(3N-2)/2},
\l{eq:microcanonicalvolume}
\ee 
where $K$ is kinetic energy, and an integration over the momenta has
been performed in the second step. The integral in
(\ref{eq:microcanonicalvolume}) behaves as $r_{ij}^{4-3N/2}$ in the
limit $r_{ij}\equiv|\vec{r}_i -\vec{r}_j| \to 0$, hence,
it diverges for $N \geq 3$, implying a diverging microcanonical entropy
(the canonical partition function also diverges). There is no way to get
rid of such a divergence other than regularizing the
Newtonian potential at short distances, by introducing, e.g., hard-core
exclusion, Pauli exclusion, etc; nevertheless, the violation of
additivity due to the long-range nature of the interaction persists in
all cases, as is represented by the occurrence of a negative specific
heat. The latter phenomenon may be heuristically justified by using the virial theorem, which for the gravitational potential
reads
\be
\langle K \rangle=-\frac{1}{2} \langle U \rangle,~ \langle K \rangle=-E,
\l{eq:virial}
\ee
where $\langle\cdot\rangle$ denotes a temporal (i.e., dynamical)
average. Since the kinetic energy $K$ is always positive, it is clear
that the virial theorem can only be valid for bound states for which
$E$ is negative. Using the equipartition theorem, we obtain the average
kinetic energy as proportional to the temperature, and, hence, Eq. (\ref{eq:virial}) implies that the specific heat
$c_V$, which is proportional to ${\rm d}E/{\rm d}T$, is negative. More
rigorously, it may be shown that regularized self-gravitating systems
confined to a box have an entropy that is a non-concave function of the
energy, see Fig. \ref{fig:convexentropy}. Since the specific heat is related to the second derivative of the entropy with respect
to energy, i.e.,
\be
\partial ^2 s/\partial \varepsilon ^2=-(c_V T^2)^{-1},
\l{eq:c_V}
\ee
it follows that in the energy range $[\varepsilon_a,\varepsilon_b]$, where the entropy
is convex, the specific heat becomes negative. For short-range additive
interactions, all states within the wider range $[\varepsilon_1,\varepsilon_2]$ would have an entropy that
is represented by the thick dashed line in
Fig.~\ref{fig:convexentropy}. In the figure, the
inverse temperature $\beta$ is also plotted as a function of $\varepsilon$,
where note that in the region of negative specific heat, the temperature
decreases as the energy increases.

\begin{figure}[ht!]
\centering
\includegraphics[width=110mm]{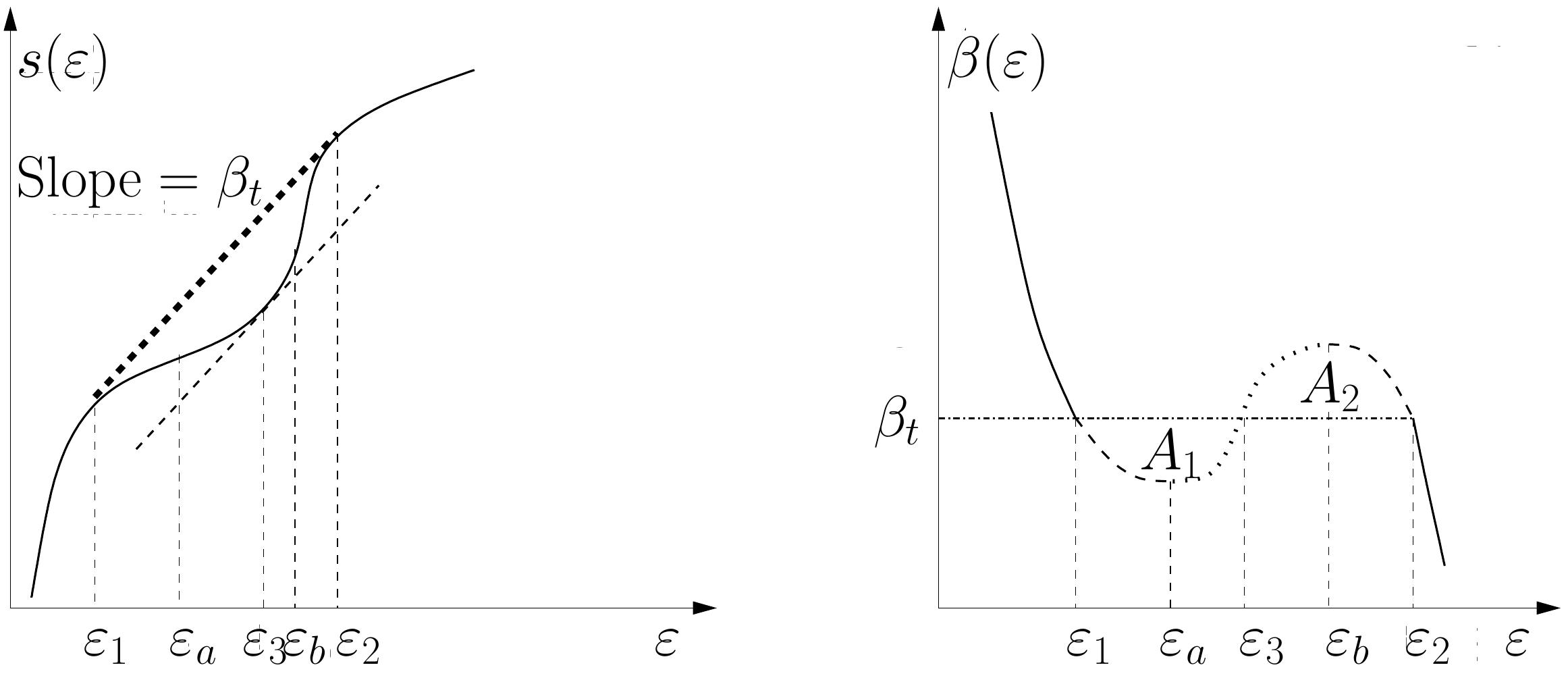}
\caption{(Left panel) Schematic shape of the microcanonical entropy per
particle as a function of the energy per particle is denoted by the solid line, which shows a ``globally"
convex region in the range $[\varepsilon_1,\varepsilon_2]$; the thick dashed line realizes 
the ``concave envelope". (Right panel) Inverse temperature $\beta$ as a function of $\varepsilon$. 
According to the Maxwell's construction, the areas $A_1$ and $A_2$ must
be equal: $A_1=A_2$. The curve $\beta(\varepsilon)$
represents states that are stable (solid line), unstable (dotted
line) and metastable (dashed lines).}
\l{fig:convexentropy}
\end{figure}

Another important example of LRI systems is that of the Euler
equations
in two dimensions, governing incompressible, inviscid fluid flow: 
\be
\frac{\partial \vec{v}}{\partial t}+(\vec{v}\cdot \vec{\nabla}) 
\vec{v}=0,~\vec{\nabla}\cdot \vec{v}=0,~\vec{v}=(v_x,v_y).
\l{eq:euler}
\ee
Using the vorticity
\be
\omega(x,y)\equiv\frac{\partial v_y}{\partial x}-\frac{\partial v_x}{\partial y},
\l{eq:vorticity}
\ee
the Euler equations may be rewritten as
\be
\frac{\partial \omega}{\partial t}+\vec{v}\cdot \vec{\nabla} \omega=0.
\l{eq:euler-again}
\ee
The long-range features of this equation may be made explicit by introducing the stream function
$\psi(x,y)$, as
\be
v_x=+\frac{\partial \psi}{\partial y},~v_y=-\frac{\partial
\psi}{\partial x},
\l{eq:stream-function}
\ee
which is related to the vorticity by the Poisson equation:
\be
\omega=-\Delta\psi.
\l{eq:stream-vorticity}
\ee
Using the Green's function $G\left(\vec{r},\vec{r'}\right)$, one may
find the solution of the Poisson equation in a given domain $D$ as
\be
\psi( \vec{r})=\int_D {\rm d}\vec{r'}~\omega(\vec{r'})\;G\left(\vec{r},\vec{r'}\right),
\l{eq:Poisson-solution}
\ee
plus surface terms. In an infinite domain, one has
\be
G\left(\vec{r},\vec{r'}\right)\equiv-\frac{1}{2
\pi} \ln |\vec{r}-\vec{r'}|.
\l{eq:Greens-function}
\ee
The energy is conserved for the Euler equation, and is given by
\bea
E&=&\int_D {\rm d}\vec{r}~ \frac{1}{2}(v_x^2+v_y^2)=\int_D {\rm
d}\vec{r}~
\frac{1}{2}\left(\nabla\psi\right)^2=\frac{1}{2}\int_D {\rm d}\vec{r}~ \omega(\vec{r})
  \psi(\vec{r})\\
  &=&-\frac{1}{4 \pi} \int_D \int_D {\rm d} \vec{r}
  {\rm d}\vec{r'}~\omega(\vec{r'})\omega(\vec{r})\ln|\vec{r}-\vec{r'}|,
\l{eq:energy-euler}  
\eea
which implies a logarithmic interaction between
vortices at distant locations, thus corresponding to a decay with
an effective exponent $\alpha=0$. For a finite
domain $D$, the Green's function contains additional surface
terms that however gives no contribution to
the energy (\ref{eq:energy-euler}) if the velocity field is tangent
to the boundary of the domain (no outflow or inflow).
One may demonstrate the non-additive features of the energy by
considering the shear flow, Fig. \ref{fig:energyEuler}, for which one has
\be
v_x=-y,~v_y=0,~\omega=1,~\psi=-y^2/2.
\l{eq:shear}
\ee
The energy per unit length, given by $E/L$, is along the $x$-direction
of the flow and within $-1 \leq y \leq 1$ larger than the energy of the
separate flows: $-1 \leq y \leq 0$, $0 \leq y \leq 1$:
\be
\frac{E}{L}=\frac{1}{3},~\frac{E_{1,2}}{L}=\frac{1}{24},
\l{eq:energy-shear}
\ee
thereby demonstrating the violation of non-additivity of the energy.

\begin{figure}[ht!]
\centering
\includegraphics[width=100mm]{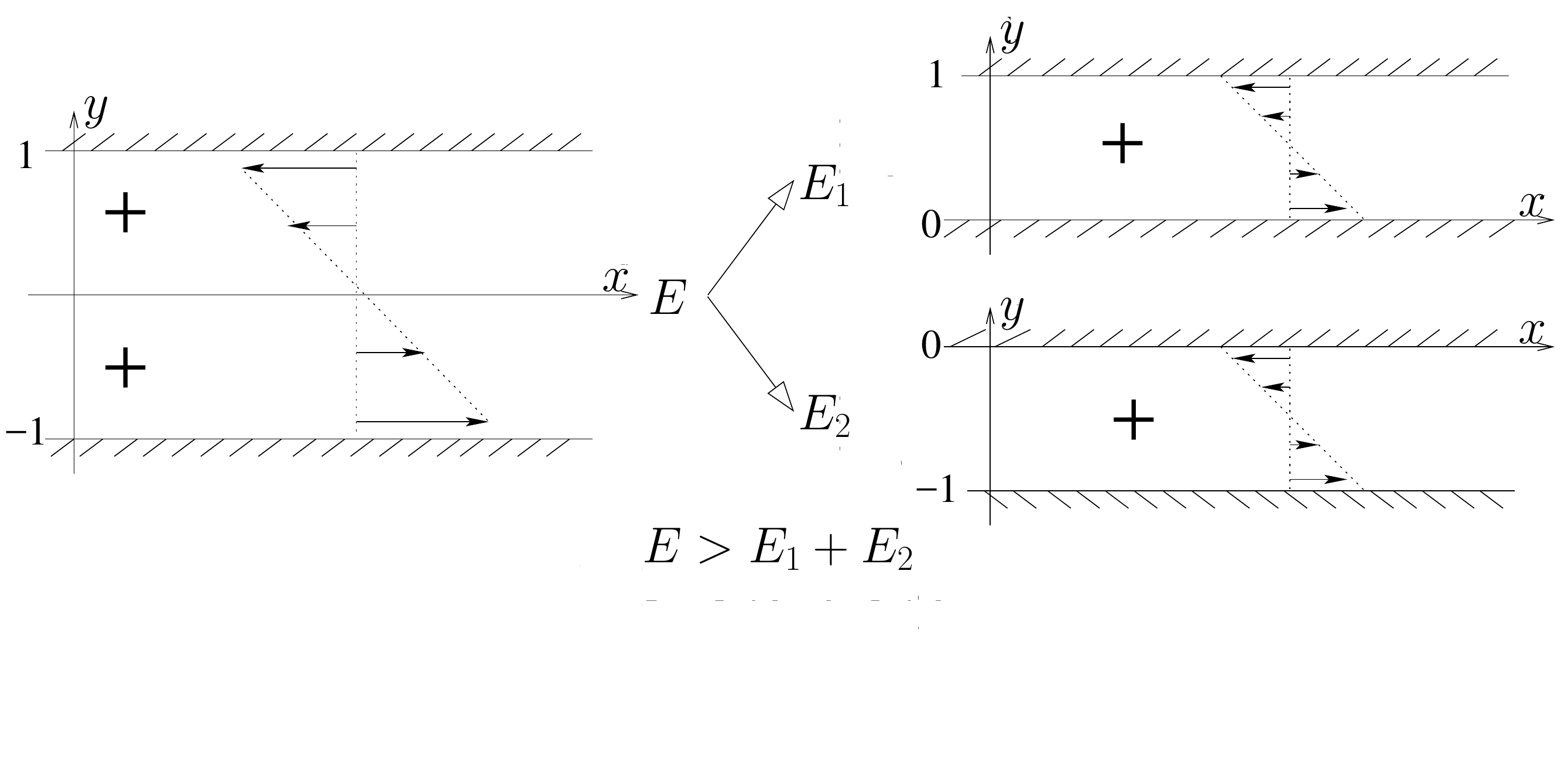}
\caption{An example showing the non-additivity of energy within the
Euler equations for
a shear flow.}
\l{fig:energyEuler}
\end{figure}

Coulomb systems constitute another relevant example of long-range interactions,
which are of type (\ref{eq:U}):
\be
U(\vec{r}_1,\ldots,\vec{r}_N)= \frac{1}{4 \pi
\varepsilon_0} \sum_{1 \le i<j \le N} e_i e_j
V(|\vec{r_i}-\vec{r_j}|),
\l{eq:U-coulomb}
\ee
where $\varepsilon_0$ is the vacuum permittivity, and $e_i$ is the charge
located at position $\vec{r_i}$. For such systems, it may be shown that the excess charge is expelled to the boundary 
of a domain, and that the bulk is neutral. A typical configuration has a distribution
of charges of equal sign surrounded by a ``cloud" of particles of opposite charge,
which ``screens" the interactions at long range. The effective two-body
potential is therefore given by
\be
V_{\rm eff} \propto \frac{\exp (-r/\lambda_D)}{r},
\l{eq:veff-coulomb}
\ee
where $\lambda_D \equiv (\varepsilon_0/(2 n e^2 \beta))^{1/2}$ is the so-called
Debye length, and $n$ is the particle density. On account of the screening,
Coulomb systems are effectively short-range. 

A plasma of electrons can be confined by a crossed electric field
$\mathbf{E}$ and a magnetic field $\mathbf{B}$ \cite{Dubin:2010}. As shown in Fig.
\ref{fig:malmberg}, the electrons are contained axially by negative voltages and radially by a uniform axial magnetic
field $B_z$. Under typical experimental conditions of density and temperature, electrons
are collisionless; They bounce axially very rapidly and drift across the
magnetic field with velocity
\be
\mathbf{v}=\frac{-\nabla \phi \times \hat{z}}{B_z}.
\ee
As is the case of an effectively incompressible fluid, the electron density $n(x,y)$ obeys the
evolution equations
\bea
&& \frac{\partial n}{\partial t}+\mathbf{v} \cdot \nabla n=0,\\
&& \Delta \phi = \frac{e n}{\varepsilon_0},
\eea 
where $-e$ is electron charge. These equations are isomorphic to the
two-dimensional Euler equations with vorticity $\omega=en/\varepsilon_0$ and stream
function $\psi=\phi/B_z$. The electron plasma may be regarded as the best 
experimental realization of two-dimensional incompressible, inviscid fluid.
\begin{figure}[ht!]
\centering
\includegraphics[width=60mm]{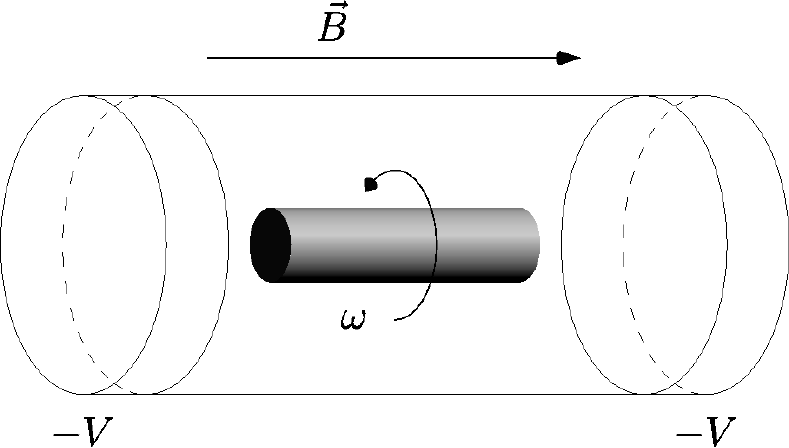}
\caption{Plasma of electrons confined by crossed electric and magnetic
fields.}
\l{fig:malmberg}
\end{figure}

Dipolar interaction is marginally long-range \cite{Bramwell:2010}: $\alpha=3$ in $d=3$. The interaction energy
of two dipoles is 
\be
E_{ij}=\frac{\mu_0}{4 \pi} \left[ \frac{\vec{\mu}_i \cdot
\vec{\mu}_j}{|\vec{r}_{ij}|^3} - \frac{ 3
(\vec{\mu}_i \cdot \vec{r}_{ij})
(\vec{\mu}_j \cdot \vec{r}_{ij})}
{|\vec{r}_{ij}|^5} \right],
\ee
where $\mu_0$ is vacuum permeability, and $\vec{\mu}_i$ is the dipolar
moment at position site $\vec{r}_i$. Because of the anisotropy of the interaction, dipolar systems are strongly frustrated:
several configurations have the same energy. For ferromagnetic samples of ellipsoidal
shape, one has the total energy
\be
E_{\rm Dipolar}= \frac{1}{2} \sum_{i,j} E_{ij} = E_0 V + \frac{1}{2}
\mu_0 \frac{(\sum_i \vec{\mu}_i)^2}{V} D,
\ee
where $E_0$ is a local-energy term that depends on the crystal
structure, and $D$ is the so-called
shape-dependent demagnetizing factor: $D=1/3$ for spherical samples, $D=0$ for needle shape 
samples, $D=1$ for disk shaped samples. The free energy of a dipolar
magnetic system is shape-independent, which implies that the macroscopic state 
cannot be ferromagnetic. However, ferromagnetism can exist in mesoscopic samples, paving the
way to the possible experimental detection of long-range effects.

An experimental apparatus where long-range forces are at play is the
free-electron laser \cite{Barre:2004}. In the linear free-electron laser, a relativistic 
electron beam propagates through a spatially  periodic magnetic field, 
interacting with the co-propagating electromagnetic wave, see Fig.
\ref{fig:fel}. Lasing occurs when the electrons bunch in a subluminar beat wave.
After scaling away the time dependence of the phenomenon, and on introducing
appropriate variables, e.g., the length $z$ along the lasing direction,
it is possible to capture the essence of the
asymptotic state by studying the following equations of motion first introduced by
Colson and Bonifacio:
\bea
\frac{{\rm d} \theta_j}{{\rm d} z} &=& p_j,\\
\frac{{\rm d} p_j}{{\rm d} z}
&=&-\mathbf{A}e^{i\theta_j}-\mathbf{A}^{\ast}e^{-i\theta_j},\\
\frac{{\rm d} \mathbf{A}}{{\rm d} z} &=&
i\delta\mathbf{A}+\frac{1}{N}\sum_j e^{-i\theta_j}.
\l{eq:felequations}
\eea
The above equations derive from the Hamiltonian
\begin{equation}
H_{\rm FEL}=\sum_{j=1}^N\frac{p_j^2}{2} -N \delta A^2 +2 A \sum_{j=1}^N
\sin(\theta_j-\varphi).
\l{eq:felhamiltonian}
\end{equation}
The $p_i$'s are related to the energies relative to the center of
mass of the $N$-electron system, and the conjugated variables $\theta_i$
characterize their positions with respect to the co-propagating
wave.  The complex electromagnetic field variable, $\mathbf{A}=A\,e^{i\varphi}$, 
defines the amplitude and the phase of the dominating
mode ($\mathbf{A}$ and $\mathbf{A}^\star$ are canonically conjugate variables).
The parameter $\delta$ measures the average deviation from the
resonance condition.

\begin{figure}[ht!]
\centering
\includegraphics[width=80mm]{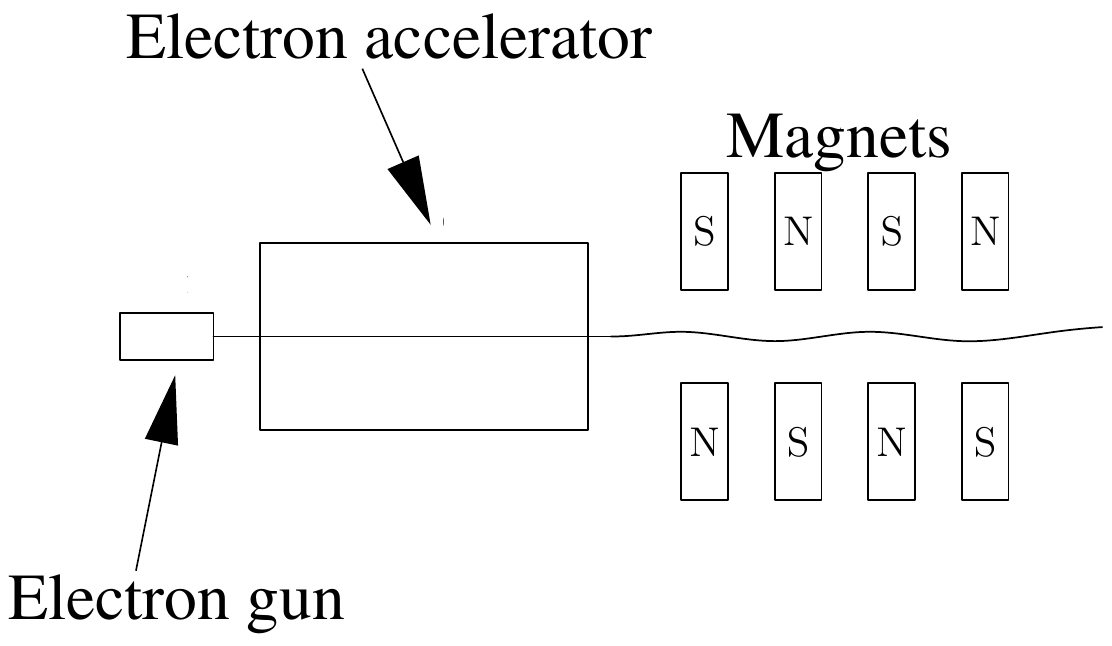}
\caption{Sketch of a linear free-electron laser.}
\l{fig:fel}
\end{figure}

\section{Dynamical evolution of LRI systems: The general scenario}
\l{sec:dynamical-evolution-generic}

In this Section, we discuss the general scenario of dynamical evolution
of isolated LRI systems. To this end, let us consider an interacting system of $N$ identical particles
of mass $m$, which we take for
simplicity of discussion to be embedded in one-dimensional ($1d$) space
(the discussions straightforwardly generalize to higher dimensions). The
Hamiltonian of the system is given by
\be
H=\sum_{i=1}^{N}\frac{p_{i}^{2}}{2m}+\sum\limits_{1 \le i<j \le N}^{N}V(|q_{i}-q_{j}|),
\l{eq:genericH}
\ee
where $q_i \in {\cal D}$ are the
canonical coordinates of the particles, and $p_i \in \mathbb{R}$ are the corresponding conjugated momenta. Here,
$V(|q_i - q_j|)$ is the two-body interaction potential between particles $i$
and $j$. We consider ${\cal D}$ to be finite and with periodic
boundary conditions, and set $m=1$ in the following without loss of generality.

A microstate of the system is specified by giving the coordinates and the
momenta of all the particles, and defines a representative point ${\bf
z}$, with $z_i\equiv (p_i,q_i);~i=1,2,\ldots,N$, in the
$2N$-dimensional phase space $\Gamma$ of the system. A distribution of
representative points in the phase space, corresponding to different
microstates of the system consistent with a given macrostate, is
characterized by the $N$-particle phase space density $\rho(z_1,z_2,\ldots,z_N,t)$,
defined such that $\rho(z_1,z_2,\ldots,z_N,t){\rm d}z_1 {\rm d}z_2\ldots {\rm d}z_N$, with
${\rm d}z_i \equiv {\rm d}p_i {\rm d}q_i$, gives
the number of representative points contained at time $t$ in an
infinitesimal volume element $\prod_{i=1}^N {\rm d}z_i$ around the point ${\bf
z}$. In view of the particles constituting the
system being identical, we consider the $N$-particle density to be symmetric in
$z_1,z_2,\ldots,z_N$. As the microstates evolve in time following the Hamilton equations
\be
\dot{p}_i=-\frac{\partial H}{\partial q_i},~~\dot{q}_i=\frac{\partial
H}{\partial p_i},
\ee
with a dot denoting derivative with respect to time, the phase space density evolves in time following the Liouville's
theorem ${\rm d}\rho/{\rm d}t=0$; Combined with the Hamilton equations, this implies
\be
\frac{\partial \rho}{\partial t}+\sum\limits_{i=1}^N \Big[p_i \frac{\partial
\rho}{\partial q_i}-\frac{\partial \rho}{\partial
p_i}\frac{\partial }
{\partial q_i}\sum\limits_{j=1,j\ne i}^{N}V(|q_{i}-q_{j}|)\Big]=0.
\l{eq:Liouville-theorem}
\ee
Since Liouville's theorem implies that $\int \prod_{i=1}^N {\rm d}z_i~\rho(z_1,z_2,\ldots,z_N,t)$ is a constant in time, we may choose this constant to be unity, implying the normalization $\int \prod_{i=1}^N {\rm d}z_i~\rho(z_1,z_2,\ldots,z_N,t)=1$.

Using the $N$-particle density $\rho(z_1,z_2,\ldots,z_N,t)$, one may define a single-particle density as
\be
f_1(z_1,t)\equiv \int \prod_{i=2}^N {\rm d}z_i~ \rho(z_1,z_2,\ldots,z_N,t).
\ee
The physical interpretation of $f_1$ is as follows: In contrast to the $2N$-dimensional $\Gamma$ space, one may construct a $2$-dimensional single-particle phase space $\mu$, with axes $(p,q)$, in which a microstate of the system is represented by $N$ representative points $(z_1,z_2,\ldots,z_N)$. A distribution of points in the $\Gamma$ space may be mapped to a distribution in the $\mu$ space. The latter is characterized by the single-particle phase space density $f_1(p,q,t)$, defined such that $f_1(p,q,t){\rm d}p{\rm d}q$ gives the number of representative points contained at time $t$ in an infinitesimal volume element 
${\rm d}p{\rm d}q$ centered at $(p,q)$. As it will turn out, it will often be convenient and meaningful to discuss the evolution of an LRI system in the $\mu$ space rather than in the higher dimensional $\Gamma$ space. Using $\rho(z_1,z_2,\ldots,z_N,t)$, one may in general define the $s$-particle density 
\be
f_s(z_1,z_2,\ldots,z_s,t)=\fr{N!}{(N-s)!}\int \prod_{i=s+1}^N {\rm d}z_i~
\rho(z_1,z_2,\ldots,z_N,t);~~s=1,2,\ldots,N.
\ee
In view of the normalization of $\rho(z_1,z_2,\ldots,z_N,t)$, one has $\int {\rm d}z~f_1(z,t)=N$, and in general, $\int \prod_{i=1}^s {\rm d}z_i~f_s(z_1,z_2,\ldots,z_s,t)=N!/(N-s)!$.

Using Eq. (\ref{eq:Liouville-theorem}), one may derive the time evolution of $f_s$ to find that a
determination of its evolution requires knowing the higher density
$f_{s+1}$. It then follows that the time evolution of the full set
$(f_1,f_2,f_3,\ldots,f_N)$ forms a coupled chain of equations, which goes
by the name of the Bogoliubov-Born-Green-Kirkwood-Yvon (BBGKY)
hierarchy. The first equation of the hierarchy reads
\be
\Big[\fr{\partial }{\partial t}+p_1\fr{\partial }{\partial
q_1}\Big]f_1(z_1,t)=\int {\rm d}z_2 ~\frac{\partial }
{\partial q_1}V(|q_{1}-q_{2}|)\frac{\partial }{\partial
p_1}f_2(z_1,z_2,t).
\l{eq:BBGKY-first}
\ee

Now, we may express the two-particle density quite generally as
\be
f_2(z_1,z_2,t)=f_1(z_1,t)f_1(z_2,t) +
g_2(z_1,z_2,t),
\l{eq:2pt-decomposition}
\ee
where $g_2$ describes the two-particle correlation. Integrating both sides with respect to $(z_1,z_2)$, it then follows that $g_2 \sim N$. 
For an LRI system, let us now invoke Kac's trick, $J \to J/N$, implying $V \to
V/N$. Then, using Eq. (\ref{eq:BBGKY-first}), and noting that $f_1 \sim N$, one obtains to leading order in $1/N$ the
evolution equation \cite{Chavanis:2008}
\be
\Big[\fr{\partial }{\partial t}+p_1\fr{\partial }{\partial
q_1}\Big]f_1(z_1,t)=\frac{\partial \Phi}
{\partial q_1}\frac{\partial }{\partial
p_1}f_1(z_1,t),
\l{eq:Vlasov}
\ee
where $\Phi[f_1](q_1,t)=\int {\rm d}z_2 ~V(|q_1-q_2|)f_1(z_2,t)$ is
the mean-field potential. Equation (\ref{eq:Vlasov}) that describes the time evolution of the single-particle phase space density is called the Vlasov equation.

In passing, we note that for a short-range system for which no Kac scaling needs to be invoked, Eq. (\ref{eq:BBGKY-first}) yields to leading order in $1/N$ the
evolution equation
\be
\Big[\fr{\partial }{\partial t}+p_1\fr{\partial }{\partial
q_1}\Big]f_1(z_1,t)=\int {\rm d}z_2 ~\frac{\partial }
{\partial q_1}V(|q_{1}-q_{2}|)\frac{\partial }{\partial
p_1}g_2(z_1,z_2,t).
\l{eq:short-range}
\ee
The difference of the above equation from the Vlasov equation
(\ref{eq:Vlasov}) is evident. To leading order in $1/N$, the time
evolution for an LRI system is governed by the mean-field potential
(besides the trivial ``streaming term" on the left hand side of both
Eqs. (\ref{eq:Vlasov}) and (\ref{eq:short-range}) that is present even
when the particles are noninteracting), with the two-particle
correlation $g_2$ providing the next higher-order correction. Instead, for a short-range system, it is the two-particle correlation that dictates the leading time-evolution of the phase space density.

Let us remark on some relevant features of the Vlasov equation
(\ref{eq:Vlasov}). The equation is evidently time-reversal invariant.
One may associate with the equation a Hamilton dynamics due to a
single-particle (mean-field) Hamiltonian ${\cal H}[f_1](p,q,t)=p^2/2+\Phi[f_1](q,t)$. Note that the presence of the mean-field potential makes this Hamiltonian a functional of the single-particle density $f_1$. One may then rewrite the Vlasov equation in terms of a Poisson bracket:
\be
\frac{\partial f_1}{\partial t}+\{f_1,{\cal H}\}=0,
\ee
which implies that Vlasov-stationary solutions ($\partial f_1/\partial
t=0$) are given by arbitrary normalizable functions of the
single-particle Hamiltonian, as $f_1(p,q)=F({\cal H})$. Another class of
stationary solutions involves those that are homogeneous in the position
coordinate ($\partial f_1/\partial q=0$), that is, $f_1(p,q)\propto
h(p)$, where $h(p)$ is a normalizable function of the momentum. Given
the extent of arbitrariness allowed in choosing the functions $F$ and
$h$, one may conclude that the Vlasov equation admits an infinite number
of stationary solutions. These stationary solutions define the so-called ``Vlasov equilibrium" state of the system. Furthermore, the equation admits an infinite number of conserved quantities, the so-called Casimirs $C[f_1]=\int {\rm d}p{\rm d}q~c(f_1(q,p))$, with $c$ an arbitrary function of its argument. In particular, the single-particle entropy $S[f_1]\equiv -\int {\rm d}p {\rm d}q~ f_1 \ln f_1$ is a conserved quantity, and is thus constant in time. 

The leading correction to the Vlasov equation is given by
\be
\Big[\fr{\partial }{\partial t}+p_1\fr{\partial }{\partial
q_1}\Big]f_1(z_1,t)-\frac{\partial \Phi}
{\partial q_1}\frac{\partial }{\partial
p_1}f_1(z_1,t)=\int {\rm d}z_2~\frac{\partial }{\partial q_1}V(|q_{1}-q_{2}|)\frac{\partial }{\partial p_1}g_2(z_1,z_2,t).
\ee
Noting that $f_1\sim N$ and $g_2 \sim N$, the above equation implies time evolution of Vlasov-stationary solutions on timescales of $O(N)$.

The second equation of the BBGKY hierarchy is
\bea
&&\Big[\fr{\partial }{\partial t}+p_1\fr{\partial }{\partial
q_1}+p_2\fr{\partial }{\partial q_2}-\frac{\partial }{\partial
q_1}V(|q_{1}-q_{2}|)\frac{\partial }{\partial
p_1}-\frac{\partial }
{\partial q_2}V(|q_{1}-q_{2}|)\frac{\partial }{\partial
p_2}\Big]f_2(z_1,z_2,t)\nonumber \\
&&=\int {\rm d}z_3~\Big[\frac{\partial }
{\partial q_1}V(|q_{1}-q_{3}|)\fr{\partial }{\partial p_1}+\frac{\partial }
{\partial q_2}V(|q_{2}-q_{3}|)\fr{\partial }{\partial p_2}\Big]f_3(z_1,z_2,z_3,t). 
\l{eq:BBGKY-second}
\eea
Similar to the decomposition (\ref{eq:2pt-decomposition}), one has for the three-point correlation 
\bea
f_3(z_1,z_2,z_3,t)&=&f_1(z_1,t)f_1(z_2,t)f_1(z_3,t)+f_1(z_1,t)g_2(z_2,z_3,t)\nonumber \\
&&+f_1(z_2,t)g_2(z_1,z_3,t)+f_1(z_3,t)g_2(z_1,z_2,t)+h(z_1,z_2,z_3,t),
\l{eq:3pt-decomposition}
\eea
where, arguing as for Eq. (\ref{eq:2pt-decomposition}), one concludes that $h \sim N$.
Using Eqs. (\ref{eq:2pt-decomposition}) and (\ref{eq:3pt-decomposition})
in Eq. (\ref{eq:BBGKY-second}), and using the Kac's scaling $V \to V/N$, one obtains 
to leading order in $1/N$ the result
\bea
\frac{\partial g_2(z_1,z_2,t)}{\partial t}&=&\Big(\frac{\partial }{\partial q_1}V(|q_1-q_2|)\frac{\partial }{\partial p_1}g_2(z_1,z_2,t)-p_1\frac{\partial }{\partial q_1}g_2(z_1,z_2,t)\nonumber \\
&&+f_1(z_1,t)\frac{\partial }{\partial q_2}\Big[\int {\rm d}z_3~V(|q_2-q_3|)f_1(z_3,t)\Big]\frac{\partial }{\partial p_2}f_1(z_2,t)\nonumber \\
&&+\frac{\partial }{\partial q_1}\Big[\int {\rm d}z_3~V(|q_1-q_3|)f_1(z_3,t)\Big]
\frac{\partial }{\partial p_1}g_2(z_1,z_2,t)\nonumber \\
&&+\frac{\partial }{\partial p_1}f_1(z_1,t)\frac{\partial }{\partial q_1}\int {\rm d}z_3~V(|q_1-q_3|)g_2(z_2,z_3,t)\Big)+\{1 \leftrightarrow
2\} \nonumber \\
&\equiv&A[f_1,g_2],
\eea
where $\{1 \leftrightarrow 2\}$ implies including terms obtained from the bracketed ones by exchanging the subscripts $1$ and $2$, and $A$ is a functional of $f_1$ and $g_2$.

On the basis of the above discussion, let us summarize the general
scenario of relaxation in LRI systems, see Fig.
\ref{fig:schematic-relaxation}. In a first stage of {\it violent
relaxation}, the system goes from a generic initial condition towards a
Vlasov-stable stationary state on a fast timescale independent of the
number of particles. In a second stage of {\it collisional relaxation},
finite-$N$ effects drive the system through a sequence of Vlasov-stable
states towards the Boltzmann-Gibbs equilibrium state on a timescale that
is strongly dependent on $N$. Often, the latter scale is a power law
$\sim N^\gamma;~~\gamma > 0$; A typical example is the Chandrasekhar
relaxation timescale for stellar systems, which is proportional to
$N/\ln N$. The Vlasov-stable stationary states have been named the
quasistationary states (QSSs), since such states emerge as the true
stationary states on
taking the $N\to \infty$ limit first, followed by the limit $t \to
\infty$. An example of a spiral galaxy ``stuck" in a QSS is shown in Fig. \ref{fig:schematic-relaxation}.

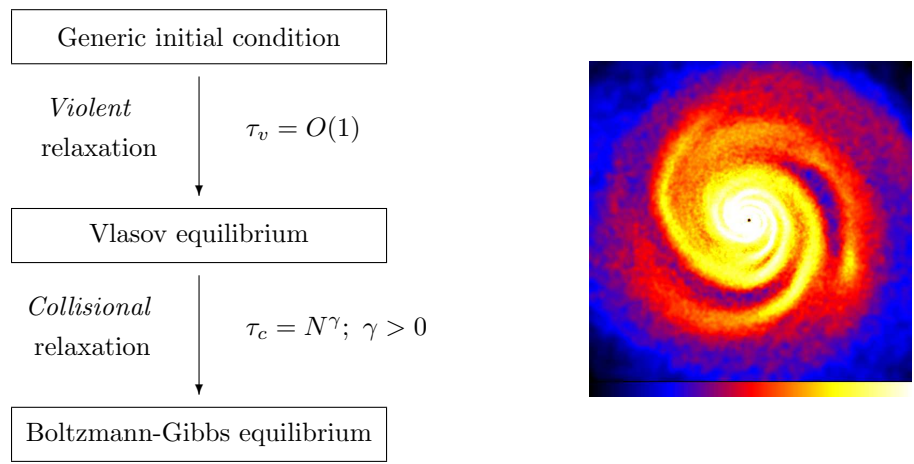
\begin{figure}[htb]
\begin{tabular}{lr}
\parbox[l]{7cm}{
{\setlength{\unitlength}{0.5pt}
\begin{picture}(200,350)(10,20)
\put(0,320){\framebox(280,40){Generic initial condition}}
\put(0,170){\framebox(280,40){Vlasov equilibrium}}
\put(0,20){\framebox(280,40){Boltzmann-Gibbs equilibrium}}
\put(175,265){$\tau_v=O(1)$} \put(175,115){$\tau_c=N^\gamma;~\gamma > 0$}
\put(25,280){{\em Violent} } \put(15,250){ relaxation}
\put(10,130){{\em Collisional}} \put(10,100){ relaxation}
\put(140,160){\vector(0,-1){90}} \put(140,310){\vector(0,-1){90}}
\end{picture}}
}&
\parbox[r]{8cm}{
\includegraphics[scale=0.2]{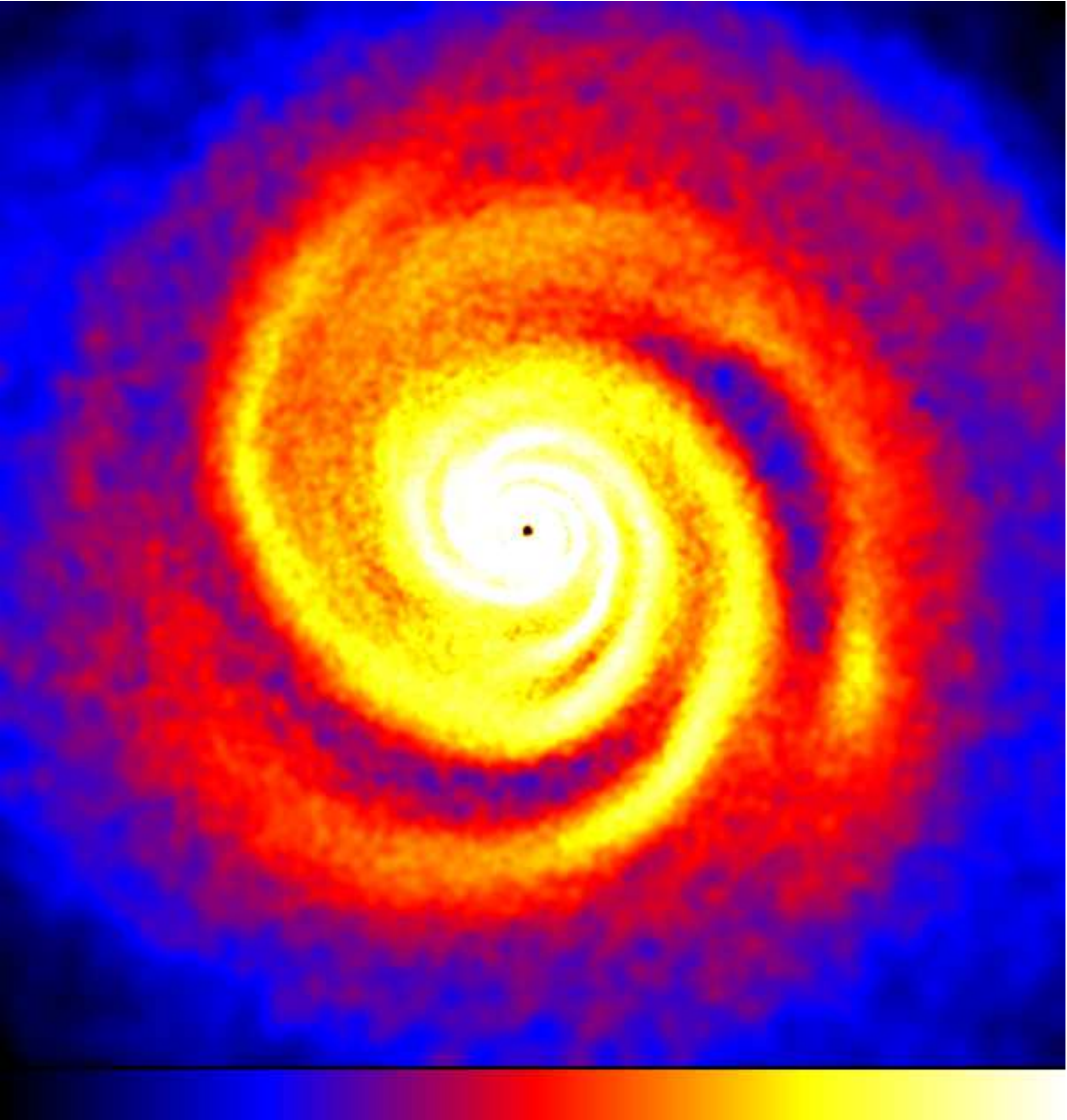}
}
\end{tabular}
\caption{(Left panel) Schematic description of the typical dynamical evolution of LRI systems. Here, $\tau_v$ and $\tau_c$ are the
{\it violent relaxation} and the {\it collisional relaxation}
timescale, respectively. (Right panel) A spiral galaxy.} 
\l{fig:schematic-relaxation}
\end{figure}

\section{A model with discrete degrees of freedom: The Kardar-Nagel
model} 
\l{sec:KN}

A solvable model of LRI systems involving discrete degrees of freedom,
which shows such features stemming from long-range interactions as
ensemble inequivalence and slow relaxation, is the so-called Kardar-Nagel
model \cite{Mukamel:2005}. The Hamiltonian reads
\be
H_{\rm KN}=-\frac{K}{2}{\sum_{i=1}^N}\left(S_iS_{i+1}-1\right)-\frac{J}{2N}
\left({\sum_{i=1}^N}S_i\right)^2,
\l{eq:modelkardar}
\ee
and involves spins $S_i=\pm 1$ occupying the sites of a one-dimensional
lattice. The spins are coupled with nearest-neighbors with strength $K$, and with
an additional Curie-Weiss ferromagnetic ($J>0$) interaction. In the
following, we set $J$ to unity without loss of generality.

\underline{The solution in the canonical ensemble:}
The canonical partition function is
\bea
Z(\beta,N)=\sum_{\{S_1,\ldots,S_N\}} e^{-\beta H}
 =\sum_{\{S_1,\ldots,S_N\}}\exp\left[{\displaystyle \frac{\beta}{2N}
 \left(\sum_{i=1}^NS_i\right)^2+\frac{\beta
 K}{2}\sum_{i=1}^N\left(S_iS_{i+1}-1\right)}\right]. \nonumber \\
 \l{eq:partfunctionNagle}
\eea
Using the Hubbard-Stratonovich
transformation, the partition function may be 
rewritten as
\bea
&&Z(\beta,N)=\sqrt{\frac{\beta N}{2\pi}}\int_{-\infty}^{\infty}{\rm d} x~e^{\displaystyle -\frac{\beta N}{2}x^2}\,
\sum_{\{S_1,\ldots,S_N\}}\Biggl[e^{\displaystyle\beta x\sum_{i=1}^N
S_i+ \frac{\beta
K}{2}\sum_{i=1}^N\left(S_iS_{i+1}-1\right)}\Biggr]\nonumber \\
&&=\sqrt{\frac{\beta N}{2\pi}}\int_{-\infty}^{\infty}{\rm
d}x~e^{\displaystyle  -N\beta \widetilde{f}\left(\beta,x\right)}.
\l{eq:partfunctionNaglebis}
\eea
The free energy may be written as
\be
\widetilde{f}(\beta,x)=\frac{1}{2}x^2 + f_0(\beta,x),
\ee
where $f_0(\beta,x)$ is the free energy of the nearest-neighbor
Ising model in an external field of strength $x$, which may be
easily derived using the transfer
matrix: $f_0(\beta,x)=-\ln(\lambda_+^N+\lambda_-^N)/(\beta N)$, where
the two eigenvalues of the transfer matrix are
\be
\lambda_\pm=e^{\beta K/2}\cosh(\beta x)\pm\sqrt{e^{\beta
K}\sinh^2(\beta x)+e^{-\beta K}}.
\ee
As $\lambda_+>\lambda_-$ for all values of $x$, only the larger
eigenvalue $\lambda_+$ is relevant in the limit
$N\rightarrow\infty$. One thus finally gets
\be
\widetilde{\phi}(\beta,x)\equiv\beta\widetilde{f}(\beta,x)= \frac{\beta }{2}x^2
-\ln \left[e^{\beta K/2}\cosh(\beta x)+\sqrt{e^{\beta
K}\sinh^2(\beta x)+e^{-\beta K}}\right].
\l{eq:philongplusshort}
\ee
In the large $N$-limit, the application of the saddle point method
to Eq. (\ref{eq:partfunctionNaglebis}) implies taking the value of $x$
that minimizes $\widetilde{\phi}(\beta,x)$ in formula
(\ref{eq:philongplusshort}), thereby yielding the free energy. From the knowledge of the free energy, one gets
either a continuous or a first-order phase transition depending on the
value of the coupling constant $K$. An expansion of
$\widetilde{f}(\beta,x)$ in powers of $x$ yields
\be
\widetilde{f}(\beta,x)= -\ln 2\cosh\frac{\beta K}{2}+ \frac{\beta
}{2}x^2\left(1-\beta  e^{\beta K}\right)+ \frac{\beta^4}{24}
e^{\beta K}\left(3e^{2\beta K}-1\right)x^4+{\cal
O}(x^6).
\l{eq:expansionlongshort}
\ee
The critical point of the continuous transition is obtained for
each $K$ by computing the value $\beta_c$ at which the quadratic
term of the expansion (\ref{eq:expansionlongshort}) vanishes, provided
the coefficient of the fourth-order term is positive, thus obtaining
$\beta_c =\exp{(-\beta_c K)}$. When also the fourth order
coefficient vanishes, i.e., for $3\exp(2\beta K)=1$, one gets the canonical
tricritical point (CTP) $K_{\rm CTP}=-\ln3/(2\sqrt{3})\simeq -0.317$.
The first-order line is obtained numerically by requiring that
$f(\beta,0)=f(\beta,x^*)$, where $x^*$ is the further local minimum
of $f$. 

\underline{The solution in the microcanonical ensemble:}
The magnetization $M \equiv \sum_{i=1}^NS_i$ may be expressed as
$M=N_+-N_-$, by introducing the number of up-spins, $N_+$, and the
number of down-spins, $N_-$. The first
term of the Hamiltonian (\ref{eq:modelkardar}) may be expressed as $-M^2/(2N)$. As two identical
neighboring spins would not contribute to the second term of the
Hamiltonian, while two different ones would give a contribution equal to
$K$, the total contribution of the second term is $KU$, where $U$ is
the number of ``kinks" in the chain, i.e., the number of links between two
neighboring spins of opposite signs.

For a chain of $N$ spins, the number of microstates corresponding to
an energy $E$ may be written as $\left(\begin{array}{ll}  N_+ \\
{\displaystyle U}/{\displaystyle 2}\end{array}\right)
\left(\begin{array}{ll}  N_- \\
{\displaystyle U}/{\displaystyle 2}\end{array}\right)$.
The formula is derived by taking into account that one has to
distribute $N_+$ spins among $U/2$ groups and $N_-$ among the
remaining $U/2$; Each of these distributions gives a binomial term,
and, since they are independent, the total number of states is the
product of the two binomials. The expression is only approximate,
because the model (\ref{eq:modelkardar}) is defined on a ring, but
nevertheless, the corrections are of order $N$, and hence, do not
contribute to the entropy. Introducing $m=M/N$, $u=U/N$ and $\varepsilon=E/N=-m^2/2+Ku$, one
thus gets the entropy as
\bea
&&\widetilde{s}(\varepsilon,m)=\frac{1}{N}\, \ln \Omega=\frac{1}{2}(1+m)\ln(1+m)+\frac{1}{2}(1-m)\ln(1-m)-u \ln
u\nonumber\\
&&-\frac{1}{2}(1+m-u)\ln(1+m-u)-\frac{1}{2}(1-m-u)\ln(1-m-u).
\l{eq:entropylongplusshort}
\eea
In the large $N$-limit, maximizing the entropy
$\widetilde{s}(\varepsilon, m)$ with respect to the magnetization $m$
leads to the final expression for the entropy: $s(\varepsilon)=\widetilde{s}(\varepsilon,
m^*)$, where $m^*$ is the equilibrium value. An expansion of
$\widetilde{s}(\varepsilon,m)$ in powers of $m$ yields
\be
\widetilde{s}(\varepsilon,m)=s_0(\varepsilon)+A_{\rm mc}m^2+B_{\rm mc}m^4+{\cal
O}(m^4),
\l{eq:expansionentropylong+short}
\ee
with the paramagnetic zero-magnetization entropy given by
\be
s_0(\varepsilon)=
-\frac{\varepsilon}{K}\ln\frac{\varepsilon}{K}-\left(1-\frac{\varepsilon}
{K}\right)\ln\left(1-\frac{\varepsilon}{K}\right),
\ee
and the expansion coefficients
\bea
A_{\rm mc}&=&
\frac{1}{2}\left[\frac{1}{K}\ln\frac{K-\varepsilon}{\varepsilon}
-\frac{\varepsilon}{K-\varepsilon}\right],\\
B_{\rm mc}&=& \frac{\varepsilon^3}{12(\varepsilon-K)^3}-\frac{K^2+K}
{4(\varepsilon-K)^2}+\frac{1}{8K\varepsilon}.
\eea
Using these expressions, it is straightforward to find the continuous transition line by requiring that $A_{\rm mc}=0$
($B_{\rm mc}<0$), finding $\beta_c=\exp(-\beta_cK)$, which is the same
equation as found in the canonical ensemble. Thus, as far as the
continuous phase transitions are concerned, the two ensembles are
equivalent. The tricritical point is obtained by the condition
$A_{\rm mc}=B_{\rm mc}=0$, giving $K_{\rm MTP}\simeq-0.359$ and
$\beta_{\rm MTP} \simeq 2.21$, which is different
from $K_{\rm CTP}\simeq -0.317$ and $\beta_{\rm CTP}=\sqrt{3}$. The
microcanonical first-order phase transition line is obtained
numerically by equating the entropies of the ferromagnetic and
paramagnetic phases. At a given transition energy, there are two
temperatures, thus leading to a temperature jump. The model also exhibits a region of negative specific heat when the phase
transition is first-order in the canonical ensemble.
The phase diagram of the model in the $(K,T)$ plane is shown in Fig.
\ref{fig:kardar}. One may observe the region of inequivalence
between the microcanonical and the canonical ensemble for $K<0$. 

\begin{figure}[ht!]
\centering
\includegraphics[width=70mm]{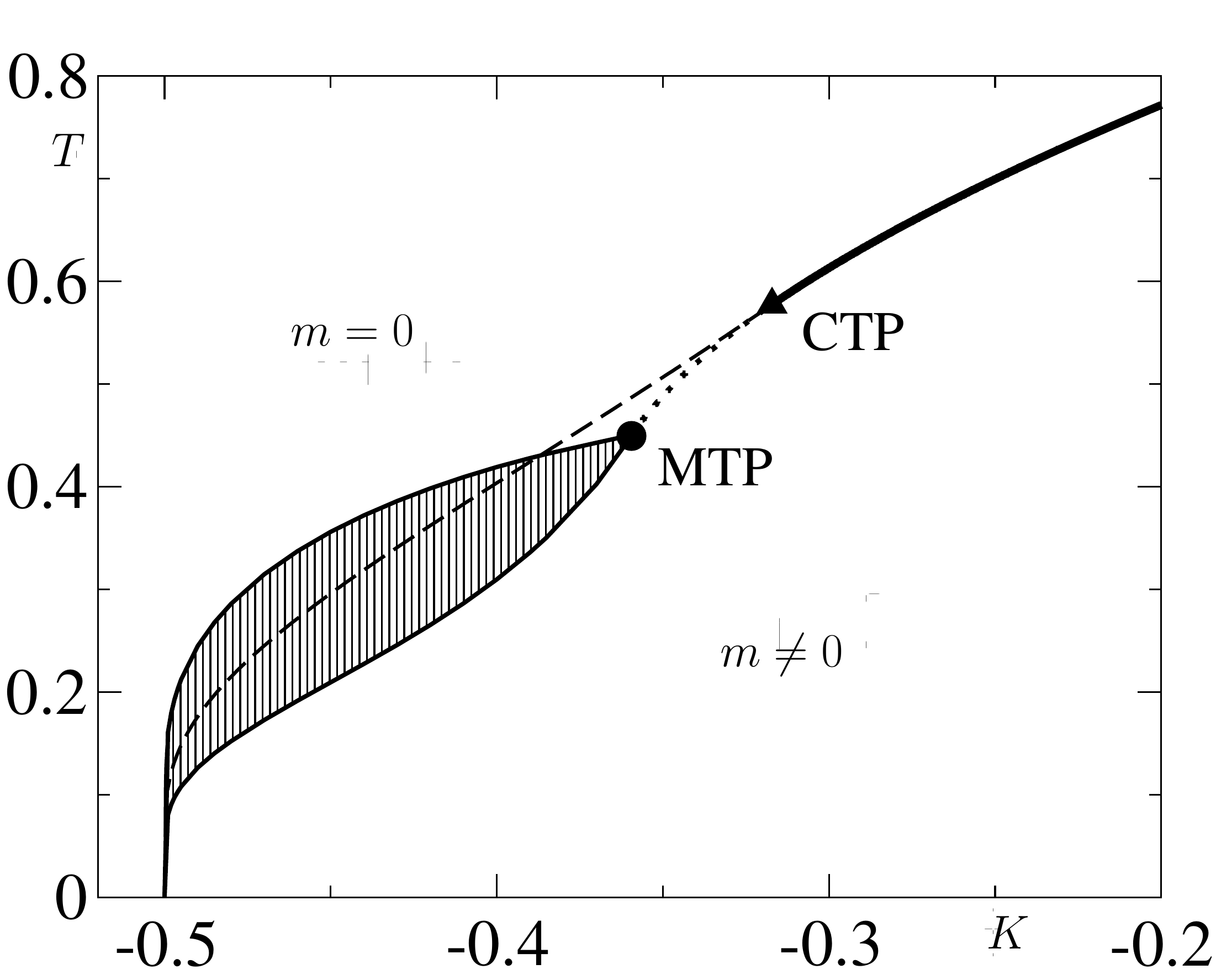}
\caption{Phase diagram of the Kardar-Nagel model, Eq.
(\ref{eq:modelkardar}). In the canonical ensemble, the large-$K$
transition is continuous (bold solid line) down to the tricritical point
CTP, where it becomes first-order (dashed line). In the microcanonical
ensemble, the continuous transition coincides with the canonical one 
at large $K$ (bold line); It persists at lower $K$ (dotted line) down to
the tricritical point MTP, where it becomes first-order, with a
branching of the transition line (solid lines). The region between these two lines (shaded area) is not accessible in the microcanonical ensemble.}
\l{fig:kardar}
\end{figure}

The Kardar-Nagel model shows broken ergodicity, because presence of
long-range interactions and the implied non-additivity make the region of macroscopic 
accessible states non-convex. Consider
positive-magnetizations states, $N_+ > N_-$, so that $0 < U < 2
N_-=N-M$, which in turn
implies in the limit $N \to \infty$ that
\be
0 \leq u= \frac{\varepsilon}{K} + \frac{J}{2K} m^2 \leq 1-m.
\l{eq:admissible}
\ee
As a consequence, the allowed magnetization-energy states are those within the
shaded area in Fig. \ref{fig:dome}. From the figure, it is evident that there are energies
(for instance, $\varepsilon=-0.35$) for which the magnetization has
three allowed values within three different intervals: one around $m=0$, and two around opposite values of $m$. Any
continuous energy-conserving dynamics initiated in one of these
intervals would now allow for a transition to states belonging to
another interval, so that ergodicity on the energy surface is broken. An
example of breaking of ergodicity is shown in Fig. \ref{fig:creutz}. In
the upper panel, the dynamics is run at an energy, $\varepsilon=-0.318$, for which the energy surface is connected and 
the system is ergodic. Nevertheless, the magnetization jumps among the three maxima of 
the entropy (shown in the inset). In the lower panel, the energy
is $\varepsilon=-0.325$, and the accessible values of the magnetization
lie in three disjoint intervals. Therefore, if the initial magnetization
lies around zero, its value remains around zero forever in time, as
shown in one of the time series. In the other, the magnetization remains at a positive value. 
No transition among the zero and the non-zero magnetization state is possible. Entropy, shown in the insets, has gaps, 
corresponding to regions where the density of states is zero.

\begin{figure}[ht!]
\centering
\includegraphics[width=90mm]{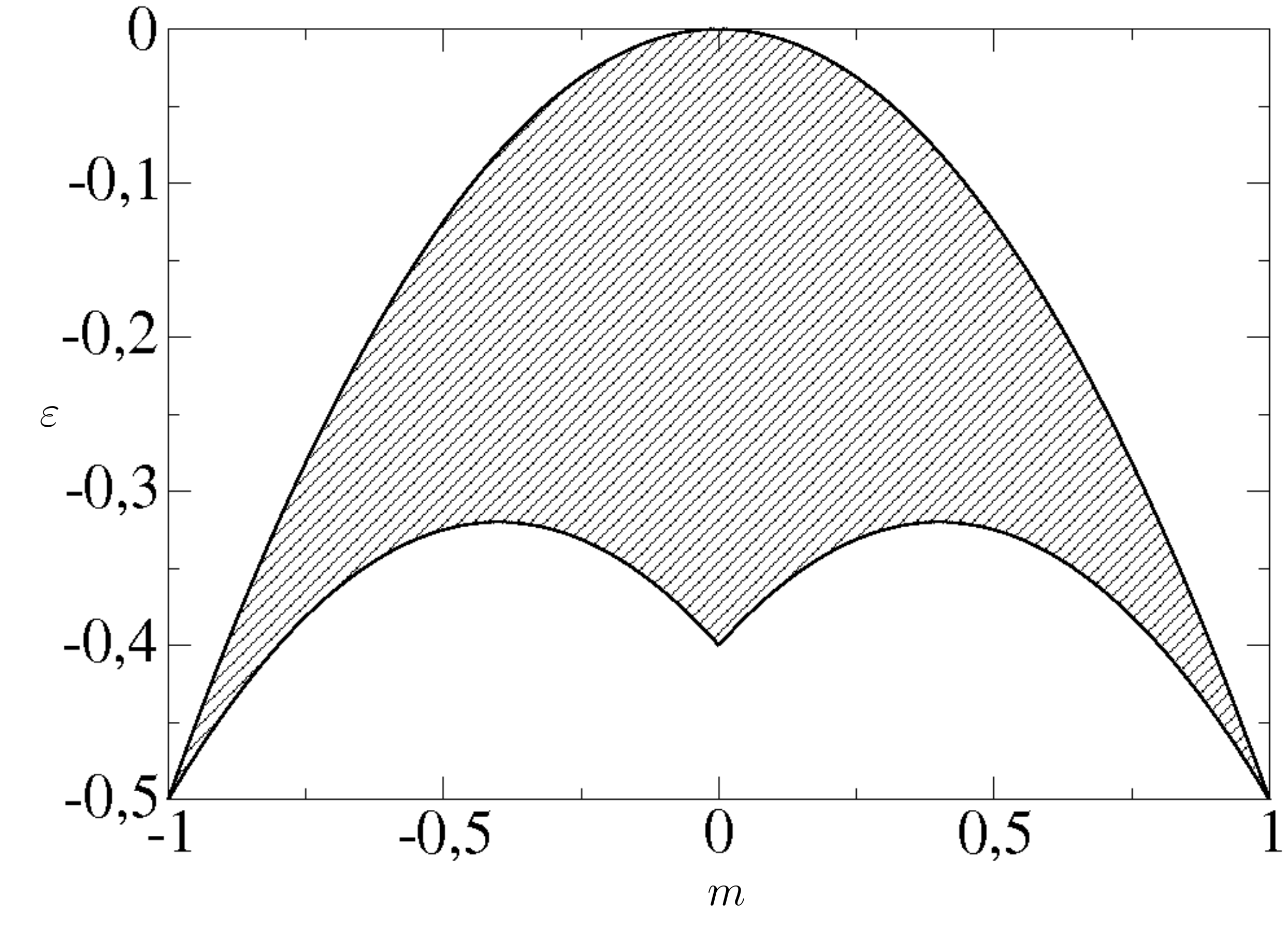}
\caption{Allowed magnetization $m$ - energy $\varepsilon$ states for the Kardar-Nagel model,
Eq. (\ref{eq:modelkardar}), with $K=-0.4$ and $J=1$.}
\l{fig:dome}
\end{figure}

\begin{figure}[ht!]
\centering
\includegraphics[width=90mm]{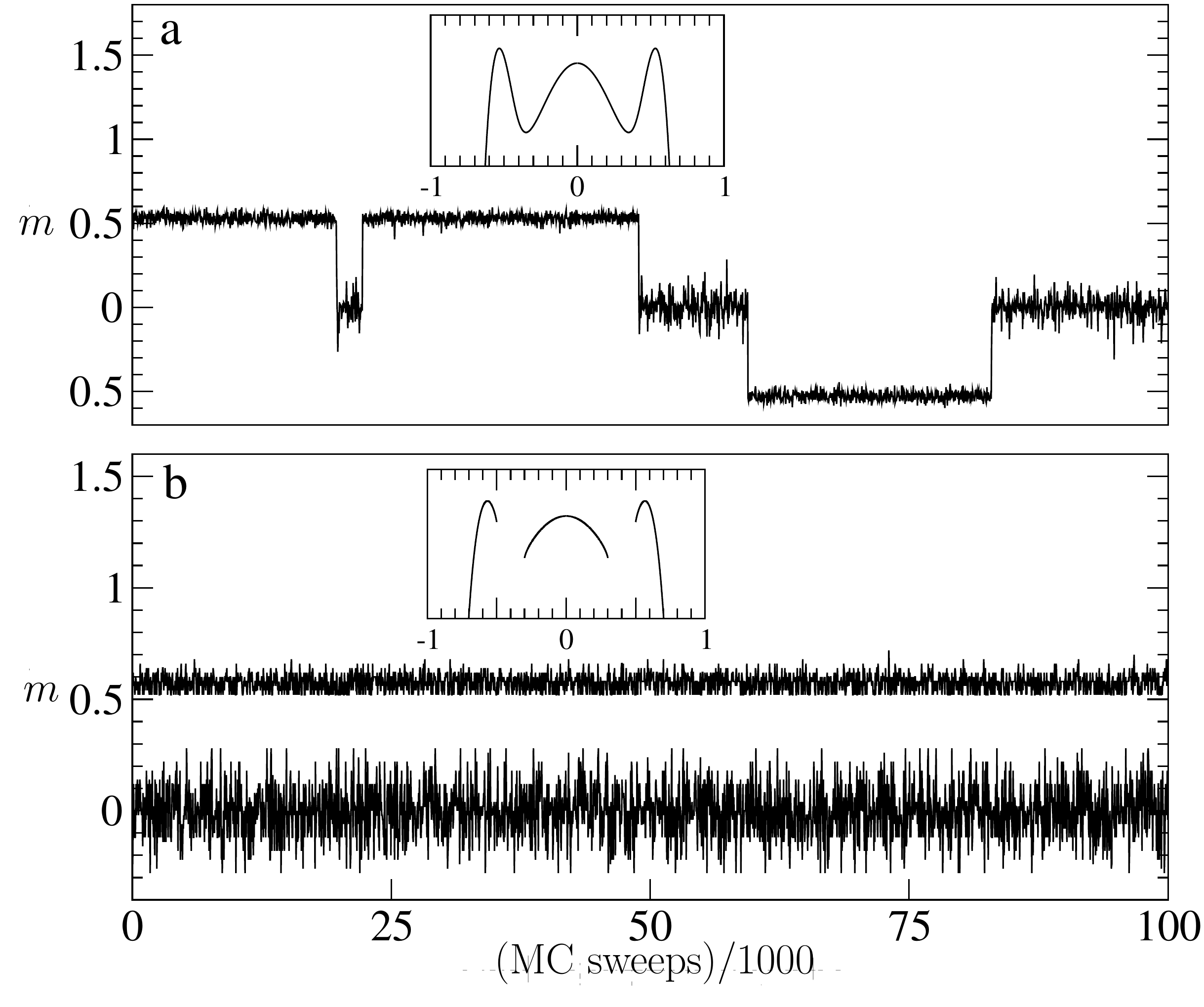}
\caption{Microcanonical Monte Carlo simulation of the Kardar-Nagel
model, Eq. {\ref{eq:modelkardar}}, with $K=-0.4$ and $J=1$ and at
different energies, showing ergodicity breaking (Lower panel).}
\l{fig:creutz}
\end{figure}

We now briefly discuss how we may simulate the dynamics of the model (\ref{eq:modelkardar})
within the microcanonical ensemble with conserved energy $E$ by Monte
Carlo simulations, using the so-called Creutz
algorithm. In this algorithm, one probes the microstates of the system
with energy $\leq E$, by adding an auxiliary variable called the ``demon", such that
\be 
E_S+E_D=E,
\l{eq:creutz-defn}
\ee
with $E_S$ being the energy of the system, and $E_D > 0$ being that of
the demon. The simulation begins with $E_S=E$, $E_D=0$, and attempt a spin flip. The move is accepted if the
energy decreases, and the excess energy resulting from the flip is given
to the demon: 
\be
E_S \to E_S - \Delta E,~E_D \to E_D+ \Delta E,~\Delta E > 0.
\l{eq:creutz-energy-distr}
\ee
If instead the energy increases due to the spin flip, the energy needed
to flip the spin is taken from the
demon:  
\be
E_S \to E_S + \Delta E,~E_D \to E_D -\Delta E,~\Delta E > 0,
\l{eq:creutz-energy-distr-again}
\ee
provided the demon has the needed energy; otherwise, the move is
rejected, but one keeps the configuration in the computation of averages. 
It can be proven that this dynamics respects detailed balance, and that the microcanonical measure 
(all configurations have equal weights on the energy surface) is stationary. 
One can also prove that the probability distribution of the demon energy
is exponential:
\be
p(E_D) \propto \exp (-\beta E_D),
\l{eq:ED-creutz}
\ee
and uses this property to determine the microcanonical inverse temperature $\beta$. 

\section{A model with continuous degrees of freedom: The Hamiltonian mean-field (HMF) model}
\l{sec:HMF}

The Hamiltonian mean-field (HMF) model is a model involving continuous
degrees of freedom and evolving
under Hamilton dynamics. The model has emerged over the years as a
prototypical model to study and elucidate the many peculiar
features resulting from long-range
interactions \cite{Antoni:1995}. The HMF model also mimics
physical systems like gravitational sheet models and free-electron
lasers. In order to derive the model, we start with the Hamiltonian
(\ref{eq:genericH}), take the mass to be unity without loss of
generality, and consider the potential to be $V(q) \propto
Jq^{-\alpha};~0\leq \alpha \leq 1$, for large $q$, so that in accordance with the Kac prescription, we scale
the coupling constant $J$ by $N$ to make the total energy extensive in $N$. Next, we assume periodic
coordinates so that boundary effects may be neglected. From now on, we denote the coordinates by
periodic variables $\theta_i$'s, with $\theta_i \in [-\pi,\pi]$, so that
$V(\theta)=V(\theta+2\pi)$. The interparticle potential $V(\theta)$, which by
definition is an even function to satisfy Newton's third law of motion, may be
expanded in a cosine Fourier series: $V(\theta)=\widetilde{v}_0/2+\sum_{k=1}^\infty
\widetilde{v}_k \cos(k\theta)$; retaining
only the first Fourier term, one obtains the HMF model. The
corresponding Hamiltonian is given by
\be
H = \sum_{i=1}^{N} \frac{p_i^2}{2} + \frac{1}{N}\sum_{1 \le i<j \le N}^N  \left[ 1 - \cos \left(\theta_i - \theta_j \right) \right],
\l{eq:hmf-H}
\ee
which effectively describes a system of globally interacting point
particles moving on a circle, with $\theta_i$ the angular
coordinate of the $i$-th particle on the circle, and $p_i$ the
corresponding conjugated momentum. In the Hamiltonian (\ref{eq:hmf-H}),
we have without loss of generality further assumed the interparticle interaction to be attractive, by
setting the coupling constant $J$ to unity. The HMF model may also be seen as a system of mean-field $XY$ spins, only that here, the
Poisson bracket of the spin components is identically zero. The Hamiltonian (\ref{eq:hmf-H}) is invariant under the O$(2)$ symmetry
group. As we show below, in thermal equilibrium and for energy densities
smaller than $\epsilon_c \equiv 3/4$, the symmetry is
spontaneously broken to result in a clustered state, thereby leading to
a continuous phase transition at $\varepsilon_c$. The order
parameter of clustering is the magnetization $m\equiv
\sqrt{m_x^2+m_y^2}$, with 
\be
(m_x,m_y) \equiv \frac{1}{N}\left( \sum_{i=1}^N \cos \theta_i, \sum_{i=1}^N \sin
\theta_i \right). 
\l{eq:hmf-mxmy}
\ee

We now derive the equilibrium solution of the model. After
the trivial Gaussian integration over the momenta, the canonical
partition function is given by
\bea
&&Z(\beta,N)=\exp
\left(-\frac{N\beta}{2}\right)\left(\frac{2\pi}{\beta}
\right)^{N/2} \nonumber \\
&&\times\int \dd \theta_1 \dots \dd \theta_N~ \exp \left\{
\frac{\beta}{2N}\left[\left(\sum_{i=1}^N \cos \theta_i \right)^2
+\left(\sum_{i=1}^N \sin \theta_i \right)^2\right]\right\}.
\l{eq:hmfcanon}
\eea
Using the Hubbard-Stratonovich transformation, we get
\bea
&&Z(\beta,N)=\exp
\left(-\frac{N\beta}{2}\right)\left(\frac{2\pi}{\beta}
\right)^{N/2}\nonumber \\
&& \times \frac{N\beta}{2\pi} \int \dd x_1 \dd x_2~ \exp
\left\{N\left[-\frac{\beta (x_1^2+x_2^2)}{2} + \ln
I_0(\beta(x_1^2+x_2^2)^{\frac{1}{2}})\right]\right\},
\l{eq:hmfhubb1}
\eea
where $I_0(z)$ is the modified Bessel function of order $0$:
$I_0(z)\equiv
\int_0^{2\pi} \dd \theta~\exp \left(z_1\cos \theta +z_2\sin \theta
\right)= \int_0^{2\pi} \dd \theta~\exp \left( z\cos \theta \right)$, with $z \equiv \left(z_1^2 + z_2^2
\right)^{1/2}$. Going to
polar coordinates in the $(x_1,x_2)$ plane yields
\be
Z(\beta,N) = \exp
\left(-\frac{N\beta}{2}\right)\left(\frac{2\pi}{\beta} \right)^{N/2}
N\beta \int_0^{\infty} \dd x~ x\exp \left\{N\left[-\frac{\beta x^2}{2}
+ \ln I_0(\beta x)\right]\right\}.
\l{eq:hmfhubb2}
\ee
In the thermodynamic limit $N\rightarrow \infty$, the integral in
(\ref{eq:hmfhubb2}) can be computed by using the saddle point method
that involves the extremization problem of finding the particular value of $x$ that
extremizes the function $\left[-\frac{\beta x^2}{2}
+ \ln I_0(\beta x)\right]$, and thus involves solving the equation
\be
x=\frac{I_1 (\beta x)}{I_0 (\beta x)},
\l{eq:solhmfcan}
\ee
where $I_1(z)=I_0'(z)$ is the modified Bessel function of order $1$. In terms of the solution of this
extremization problem, one finally obtains the rescaled free energy per particle as
\be
\phi(\beta)\equiv\beta f(\beta)= \frac{\beta}{2} -\frac{1}{2}\ln 2\pi
+\frac{1}{2}\ln \beta +\inf_{x\ge 0} \left[\frac{\beta x^2}{2}-\ln
I_0(\beta x)\right],
\l{eq:freehmf}
\ee
where note that one has to choose the particular solution of
(\ref{eq:solhmfcan}) that minimizes the free energy (\ref{eq:freehmf}).
For $\beta \leq 2$, the solution of Eq. (\ref{eq:solhmfcan}) is given by
$x=m^*=0$, while for $\beta \geq 2$, the solution
monotonically increases with $\beta$, approaching $m^*=1$ for $\beta
\rightarrow \infty$. The solution $m^*=0$ of (\ref{eq:solhmfcan}),
present for all values of $\beta$, may be discarded for $\beta >
2$, since it does not minimize the free energy.
One may show that the value $m^*$ realizing the
extremum in Eq. (\ref{eq:freehmf}) is equal to the spontaneous
magnetization in equilibrium. Note from the foregoing discussions that
the spontaneous magnetization is defined only up to its modulus, while there is a
continuous degeneracy in its direction. We have thus shown that the HMF
model displays a continuous phase transition at $\beta_c=2$ ($T_c=0.5$). The derivative of
the rescaled free energy with respect to $\beta$ gives the energy
per particle as
\be
\l{eq:enerhmfcan} \veps (\beta) = \frac{1}{2\beta} + \frac{1}{2}
-\frac{1}{2}(m^*(\beta))^2.
\ee
As already evident from the Hamiltonian, the lower bound of $\veps$
is $0$. At the critical temperature, the energy is
$\veps_c=3/4$. Since we have shown that the HMF model has a
continuous phase
transition in the canonical ensemble, we conclude that microcanonical
and canonical ensembles are equivalent for this model.

For the system (\ref{eq:hmf-H}), one may easily write down the following Vlasov equation for the evolution of the single-particle
phase space density $f(q,p,t)$ (cf. Eq. (\ref{eq:Vlasov})):
\be
\frac{\partial f}{\partial t}+p\frac{\partial f}{\partial
\theta} -\frac{\partial \Phi}{\partial \theta}
\frac{\partial f}{\partial p}=0,
\l{eq:Vlasov-hmf}
\ee
where one has the mean-field potential
\be
\Phi[f](\theta,t)=-\int_0^{2\pi}{\rm
d}\theta'\int_{-\infty}^{\infty}{\rm d}
p~\cos(\theta-\theta')f(\theta',p,t).
\l{eq:Phi-hmf}
\ee
For distributions that are homogeneous with respect to $\theta$, the
mean-field potential evaluates to zero, implying that such distributions
are stationary solutions of the Vlasov equation (\ref{eq:Vlasov-hmf}).
Let us denote such homogeneous stationary solutions by $f_0(p)$. Since
stationarity does not guarantee stability, one may study the stability
of such homogeneous distributions with respect to small perturbations, by
linearizing the Vlasov equation (\ref{eq:Vlasov-hmf}) about $f_0(p)$.
One obtains the result that the homogeneous distribution $f_0(p)$ is stable if
and only if the quantity 
\be 
I \equiv
1+\frac{1}{2}\int_{-\infty}^{\infty}{\rm d}p~\frac{f_0'(p)}{p}
\l{eq:therholdcon}
\ee
is
positive. Such a condition reveals that there can be an
infinite number of Vlasov-stable stationary distributions. Let us briefly
discuss some examples of $f_0(p)$.

\begin{itemize}
\item The first one is the Gaussian distribution: $f_0(p)\sim
\exp({-\beta p^2/2})$, which is expected at equilibrium. With the
threshold condition (\ref{eq:therholdcon}), one recovers the result due
to statistical mechanics reviewed above that the critical inverse
temperature is $\beta_c=2$, and its associated critical stability
threshold is $\varepsilon^\star=\varepsilon_c=3/4$. 

\item The second example is the so-called water-bag distribution,
which has often been used in the past
to numerically demonstrate the out-of-equilibrium properties of the HMF model.
Such a distribution comprises momentum uniformly distributed in a given interval
$[-p_0,p_0]$, where the parameter $p_0$ is related to the energy density
as $p_0=\sqrt{6\varepsilon-3}$. In this case, one obtains the critical
stability threshold as $\epsilon^\star=7/12$: the state is linearly
stable for energies larger than $\epsilon^\star$, and is unstable below.
\end{itemize}
Let us emphasize that the above examples are Vlasov-stable stationary
solutions that are possible among infinitely many others, and there is no reason to emphasize
one over the other. 

The existence of infinitely many Vlasov-stable stationary solutions of
the HMF model implies that when starting initially from one such
solution in the stable regime (e.g., the water-bag distribution at energy
density $\varepsilon > \varepsilon^\star=7/12$), and evolving under the
Hamilton equations derived from the Hamiltonian (\ref{eq:hmf-H}):
\be
\fr{\dd \th_i}{\dd t}=p_i, ~\fr{\dd p_i}{\dd t}=-m_x\sin \th_i+m_y\cos \th_i, 
\l{eq:hameq}
\ee
the magnetization in an
infinite system should remain zero at all times. For finite $N$, however, finite-$N$ effects drive the system away
from the water-bag state, and through other stable stationary states. Such a slow
quasi-stationary evolution across an infinite number of Vlasov-stable
stationary states (note: stationary only in the limit $N\to \infty$) ends with the system in the
Boltzmann-Gibbs (BG) equilibrium state, see Fig. \ref{fig:figdiffN}. For the
HMF model, it has been rigorously proven that Vlasov-stable homogeneous
distributions do not evolve on timescales of order smaller or equal to
$N$. Indeed, a scaling $\sim N^\gamma;~\gamma>0$, for the timescale of 
relaxation towards the BG equilibrium state has been observed in
simulations \cite{Yamaguchi:2004}. 
At long times, one has $m(t) \sim (1/\sqrt{N})e^{t/N^\gamma}$ for $t \gg
N^\gamma$, where the prefactor accounts for finite-$N$ fluctuations. For $\eps < \eps^*$, linear instability results in a faster
relaxation towards equilibrium as $m(t) \sim (1/\sqrt{N})e^{\gamma t}$
for $t \gg 1/\gamma$. Here, $\gamma^2=6\left(7/12-\eps\right)$ is
independent of $N$. Thus, there are no QSSs for energies below $\eps^*$. Note
that the slow relaxation to BG equilibrium depicted in Fig. \ref{fig:figdiffN} is
consistent with the general scenario shown in Fig.
\ref{fig:schematic-relaxation}.
Recently, a theory using the so-called core-halo distributions has been
proposed to quantitatively predict the properties of the QSSs \cite{Levin:2014}.
A non-mean-field version of the HMF model, the so-called $\alpha$-HMF model,
has been proposed and studied in Ref. \cite{Anteneodo:1998} in the context of Lyapunov
exponents, and in discussing the dominance of the mean-field mode in
dictating the dynamics \cite{Gupta:2012}. 

\begin{figure}[ht]
\begin{center}
\includegraphics[width=80mm]{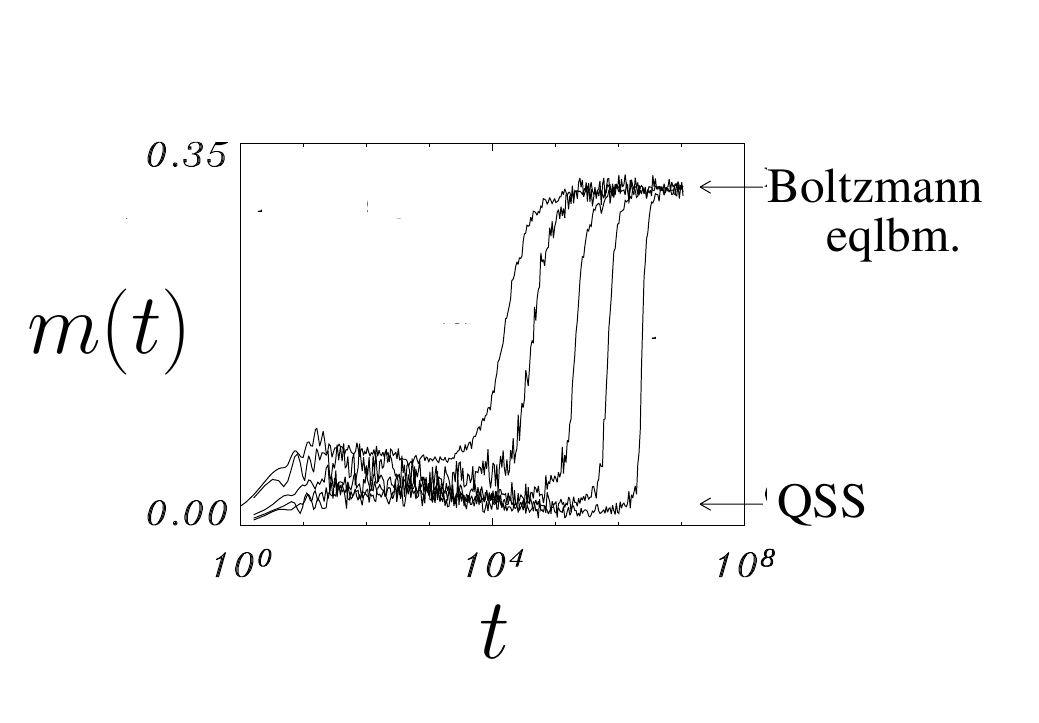}
\end{center} \caption{Time evolution of the modulus of the
magnetization $m(t)$ for different particle numbers in the HMF model
(\ref{eq:hmf-H}): $N=10^3$,
$2\times10^3$, $5\times10^3$, $10^4$ and $2\times10^4$ from left to right. The energy
density is $\varepsilon=0.69$. The values
of the magnetization indicated by the horizontal arrows refer respectively to the value expected in equilibrium (labelled BG) and the
one corresponding to a homogeneous QSS.}
\l{fig:figdiffN}
\end{figure}

\subsection{An experimental realization of the HMF model: Atoms in optical cavities}
\l{sec:Morigi}

Atoms interacting with a single-mode standing electromagnetic wave due
to light trapped in a high-finesse optical cavity are subject to an
inter-particle interaction that is long-ranged owing to multiple
coherent scattering of photons by the atoms into the wave mode
\cite{Schutz:2014,Schutz:2015,Jager:2016}. The set-up is shown in Fig. \ref{fig:morigi-setup}, which also
shows optical pumping by a transverse laser of intensity $\Omega^2$ to
counter the inevitable cavity losses quantified by the cavity linewidth
$\kappa$. As regards the interaction with the electromagnetic
field, each atom may be regarded as a two-level system, where the
transition frequency between the two levels is $\omega_0$. Considering
$N$ identical atoms of mass $m$ confined in one dimension along the
cavity axis (taken to be the $x$-axis), and denoting the standing wave
with wave number $k$ by $\cos(kx)$, the sum of the electric-field
amplitudes coherently scattered by the atoms at time $t$ depends on
their instantaneous positions $x_1,\ldots,x_N$, and is proportional to the quantity 
$\Theta\equiv \sum_{j=1}^N\cos(kx_j)/N$,
so that the cavity electric field at time $t$ is $E(t)\propto
\sqrt{N\overline{n}}\Theta$. Here, $\overline{n}$ is
the maximum intra-cavity photon number per atom, given by
$\overline{n}\equiv N\Omega^2\alpha^2/(\kappa^2+\Delta_c^2)$,  with
$\alpha \equiv g/\Delta_a$ being the ratio between the cavity vacuum
Rabi frequency and the detuning $\Delta_a\equiv \omega_L-\omega_0$
between the laser and the atomic transition frequency, and $\Delta_c\equiv
\omega_L-\omega_c$ being the detuning between the laser and the
cavity-mode frequency. The quantity
$\Theta$ characterizes the amount of spatial ordering of atoms within
the cavity mode, with $\Theta=0$ corresponding to atoms being uniformly
distributed and the resulting vanishing of the cavity field, and
$|\Theta| \ne 0$ implying spatial ordering. The wave number $k$ is related to the linear dimension $L$ of the cavity
through $k=2\pi/\lambda$ and $L=q\lambda$, where $\lambda$ is the
wavelength of the standing wave, and $q \in \mathbb{N}$.

\begin{figure}[!ht]
\centering
\includegraphics[width=2.5in]{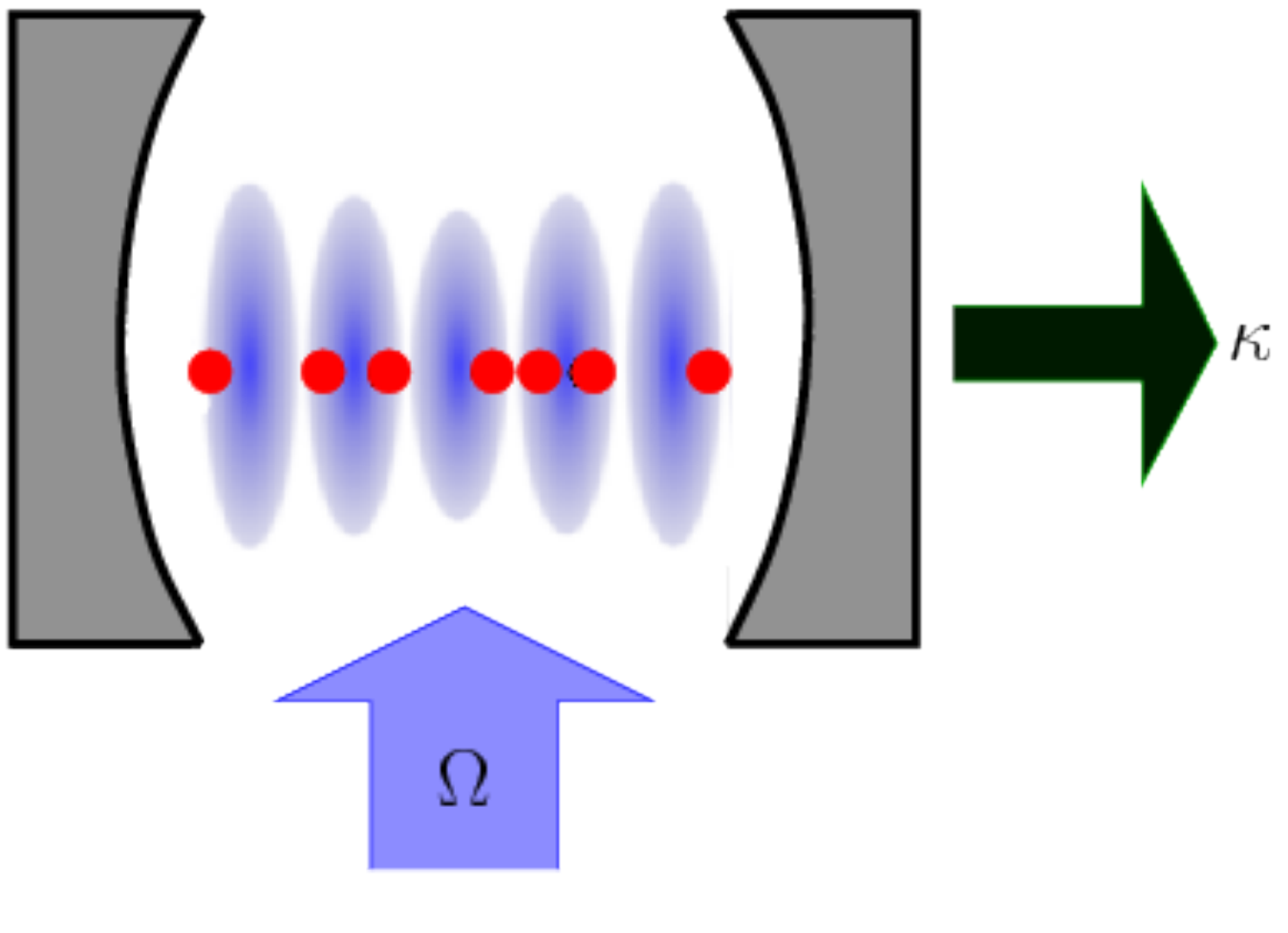}
\caption{Atoms interacting with a single-mode standing electromagnetic wave in a cavity
of linewidth $\kappa$, and being driven by a transverse laser with
intensity $\Omega^2$.}
\l{fig:morigi-setup}
\end{figure}

The dynamics of the system is studied by analyzing the
time evolution of the $N$-atom phase space distribution
$f_N(x_1,\ldots,x_N, p_1,\ldots,p_N,t)$ at time $t$, with $p_j$'s
denoting the momenta conjugate to the positions $x_j$. Treating the
cavity field quantum mechanically, and regarding the atoms as
classically polarizable particles with semi-classical center-of-mass
dynamics, it may be shown that the distribution $f_N$ evolves in time according to the Fokker-Planck equation (FPE) \cite{Schutz:2014,Schutz:2015}
\be
\partial_t f_N+\{f_N,H\}=-\overline{n}\Gamma\sum_{i=1}^N
\sin(kx_i)\sum_{j=1}^N
\partial_{p_i}\sin(kx_j)\Big(p_j+\frac{m}{\beta}\partial_{p_j}\Big)f_N.
\l{eq:atom-FPE}
\ee
Here, $\Gamma \equiv 8\omega_r \kappa \Delta_c/(\Delta_c^2+\kappa^2)$,
$\hbar\beta\equiv-4\Delta_c/(\Delta_c^2+\kappa^2)$, $\hbar$ is the
reduced Planck constant, $\omega_r \equiv \hbar k^2/(2m)$ is the
recoil frequency due to collision between an atom and a photon, while the Hamiltonian $H$ is given by
\be
H \equiv \sum_{j=1}^{N}\frac{p_{j}^{2}}{2}-NJ\Theta^2;~~J \equiv -\hbar \Delta_c \overline{n}.
\l{eq:atom-H}
\ee
The semi-classical limit is valid under the condition of $\kappa$ being
larger than $\omega_r$, while Eq. (\ref{eq:atom-FPE}) holds in a parameter
regime in which processes describing a virtual scattering of cavity
photons, which scale with the dynamical Stark shift of the cavity field
$U=g\alpha$, are negligible. The Hamiltonian $H$
describes the conservative dynamical evolution of $f_N$ in the limit of
vanishing cavity losses (or for times sufficiently small such that
dissipative effects are negligible), and contains the photon-mediated long-ranged
(mean-field) interaction between the atoms encoded in the second term on
the right hand side (rhs) of Eq. (\ref{eq:atom-H}). Note that the
interaction is attractive (respectively, repulsive) when $\Delta_c$ is
negative (respectively, positive). Cavity losses lead to damping and
diffusion, which is described by the rhs of Eq. (\ref{eq:atom-FPE}).

Let us now consider the case of effective attractive interactions
between the atoms and the cavity field ($\Delta_c < 0$), and study the
dynamics of the system in the limit in which the effect of the
dissipation may be neglected, that is, for sufficiently small times. In
this limit, the dynamics of the $N$ atoms is conservative and governed
by the Hamiltonian (\ref{eq:atom-H}). The positions $x_j$ of the atoms
enter the Hamiltonian only as $kx_j$, so that we may define the phase
variables $\theta_j \equiv k x_j = 2\pi x_j/\lambda$ for $j =
1,\ldots,N$. Using $L=q\lambda$, and setting the origin of the
$x$-axis in the center of the cavity, we have $x_j \in
[-q\lambda/2,q\lambda/2]$, so that on using the periodicity of the
cosine function, we can take the phase variables $\theta_j$ modulo $q$,
yielding $\theta_j \in [-\pi,\pi]$. Then, by measuring lengths in units
of the reciprocal wavenumber $k^{-1} = \lambda/(2\pi)$ of the cavity
standing wave, masses in units of the mass of the atoms $m$, and
energies in units of $\hbar\Delta_c$, the Hamiltonian may be rewritten in dimensionless form as
\be
H =  \sum_{j=1}^{N}\frac{(p_\theta)_j^{2}}{2} - \overline{n} N \Theta^2\,,
\l{eq:atom-H-dl}
\ee
where, in terms of the $\theta$ variables, $\Theta$ is now expressed as
$\Theta = \sum_{j = 1}^N \cos\theta_j/N$.
The $(p_\theta)_j$'s are 
the momenta canonically conjugated to the $\theta_j$ variables.

The similarity between the system with Hamiltonian (\ref{eq:atom-H-dl}) and
the HMF model is now well
apparent. Hence, the dynamics of a system of atoms interacting with
light in a cavity in the dissipationless limit is equivalent to that of
a model that differs from the HMF model in zero field just
for the fact that particles in the former interact only with the
$x$-component of the magnetization. 

\section{Ubiquity of the quasistationary behavior under different
energy-conserving dynamics} 
\subsection{HMF model in presence of three-body collisions}
\l{sec:three-body}

In this Section, we address the question of robustness of QSSs with
respect to stochastic dynamics of an isolated system within a
microcanonical ensemble. To this end, we generalize the HMF model to
include stochastic dynamical moves in addition to the deterministic
ones, Eq. (\ref{eq:hameq}). The generalized HMF model follows a
piecewise deterministic dynamics, whereby the Hamiltonian evolution is
randomly interrupted by stochastic interparticle collisions that
conserve energy and momentum \cite{Gupta:2010-1,Gupta:2010-2}. We consider collisions in which only the
momenta are updated stochastically. Since the momentum variable in the
HMF model is one-dimensional, and there are two conservation laws for
the momentum and the energy, one has to resort to three-particle
collisions. Namely, three random particles, $(i,j,k)$, collide and their
momenta are updated stochastically, $(p_i,p_j,p_k) \rightarrow
(q_i,q_j,q_k)$, while conserving energy and momentum and keeping the
phases unchanged. Thus, the model evolves under the following repetitive
sequence of events: deterministic evolution, Eq. (\ref{eq:hameq}), for a time interval whose length is exponentially distributed, followed by a single instantaneous sweep of the system for three-particle collisions, which consists of $N^3$ collision attempts. 

In presence of collisions, in order to discuss the evolution of the
single-particle phase space density in the limit $N \rightarrow \infty$, we need to consider instead of the
Vlasov equation the appropriate Boltzmann equation that takes into
account the collisional dynamics. The equation is given by 
\bea
&&\fr{\p f}{\p t}+p\fr{\p f}{\p \th}-\fr{\p \Phi}{\p
\th}\fr{\p f}{\p p}=\left(\fr{\p f}{\p t}\right)_c,\l{eq:Boltz}\\
&&\left(\fr{\p f}{\p t}\right)_c=\int {\rm d}\eta
R[f(\th,q,t)f(\th',q',t)f(\th'',q'',t)-f(\th,p,t)f(\th',p',t)f(\th'',p'',t)],
\nonumber \\
\l{eq:collisionterm} \\
&&R=\alpha\delta(p+p'+p''-q-q'-q'')\delta\left(\fr{1}{2}(p^2+p'^2+p''^2)-\fr{1}{2}(q^2+q'^2+q''^2)\right),
\eea
where we have ${\rm d}\eta \equiv {\rm d}p'{\rm d}p''{\rm d}q{\rm d}q'{\rm
d}q''{\rm d}\th'{\rm d}\th''$. Equation (\ref{eq:collisionterm})
represents the three-body collision term, and $R$ is the rate for
collisions $(p,p',p'') \rightarrow (q,q',q'')$ that conserve energy and
momentum. The constant $\alpha$ has the dimension of 1/(time) and sets
the scale for collisions: On average, there is one collision after every
time interval $\alpha^{-1}$. We refer to the Boltzmann equation with
$\alpha=0$ as the Vlasov-equation limit. Note that both the Boltzmann
and the Vlasov equation are valid for infinite $N$, and have
size-dependent correction terms when $N$ is finite. In the Vlasov limit,
any state that is homogeneous in angles but with an arbitrary momentum
distribution is stationary; as discussed in Section \ref{sec:HMF}, in this limit, the QSSs
are related to the linear stability of the stationary solutions chosen
as the initial state. Recall, for example, that the so-called water-bag state is linearly stable for
energies in the range $\eps^* \equiv 7/12 < \eps < \eps_c$, when it
manifests as a QSS. The water-bag state may be realized by sampling
independently the angles uniformly in $[-\pi,\pi]$ and the momenta
uniformly in $[-p_0,p_0]$, with $p_0=\sqrt{6\varepsilon-3}$. 

Let us now turn to a discussion of QSSs in the generalized HMF model, i.e., under noisy microcanonical evolution, in the light of the Boltzmann equation. First, we note that unlike the Vlasov equation, a homogeneous state with an arbitrary momentum distribution is not stationary under the Boltzmann equation; instead, only a Gaussian distribution is stationary. Suppose we start with an initial homogeneous state with uniformly distributed momenta. Then, under the dynamics, the momentum distribution will evolve towards the stationary Gaussian distribution. Interestingly, although the momentum distribution evolves, the initial $\th$ distribution does not change in time, since for homogeneous $\th$ distribution, the $p$ and $\th$ distributions evolve independently. In a finite system, however, there are fluctuations in the initial state. These fluctuations make the homogeneous state with Gaussian-distributed momenta linearly unstable under the Boltzmann equation at all energies $\eps < \eps_c$, as we demonstrate below. This results in a fast relaxation towards equilibrium.

One may study the linear instability of a homogeneous state with
Gaussian-distributed momenta at energies below $\eps_c$ and under the evolution
given by the Boltzmann equation. We now summarize the essential steps, for the simple case of energies just below the critical point. The stability analysis is carried out by linearizing Eq.
(\ref{eq:Boltz}) about the homogeneous state. We expand $f(\th,p,t)$ as
$f(\th,p,t)=f^{(0)}(p)[1+\lambda f^{(1)}(\th,p,t)]$ with
$f^{(0)}(p)=e^{-p^2/2T}/(2\pi\sqrt{2\pi T})$. Here, since the initial
angles and momenta are sampled independently according to $f^{(0)}(p)$,
fluctuations for finite $N$ make the small parameter $\lambda$ of
$O(1/\sqrt{N})$. At long times, the dynamics is dominated by the
eigenmode with the largest eigenvalue of the linearized Boltzmann
equation, so that $f^{(1)}(\th,p,t)=f^{(1)}_{k}(p,\o)e^{i(k\th+\o t)}$.
Since the mean-field potential $\Phi$ in Eq. (\ref{eq:Boltz}) involves $e^{\pm i\th}$, one needs to consider only $k=\pm 1$. The coefficients $f^{(1)}_{\pm 1}$ then satisfy
\bea
&&\!\!\!\!\!\!\!\!\pm ipf^{(1)}_{\pm 1}(p,\o)\mp \fr{2\pi}{2if^{(0)}}\fr{\p 
         f^{(0)}}{\p p}\int {\rm d}p'f^{(0)}(p')f^{(1)}_{\pm 1}(p',\o)\nonumber \\
         &&\!\!\!\!\!\!\!\!+(4\pi)^2\int
         {\rm d}p'{\rm d}p''{\rm d}q{\rm d}q'{\rm d}q''Rf^{(0)}(p')f^{(0)}(p'')\nonumber \\
         &&\!\!\!\!\!\!\!\!\times [f^{(1)}_{\pm 1}(p,\o)-f^{(1)}_{\pm 1}(q,\o)]=-i\o f^{(1)}_{\pm 1}(p,\o).
\l{lin2}
\eea
Treating $\alpha$ as a small parameter, we solve the above equation
perturbatively in $\alpha$. In the absence of collisions ($\alpha=0$),
the above analysis reduces to that of the Vlasov equation and to the
unperturbed solutions, namely, the frequencies $\o^{(0)}$ and the
coefficients $f^{(1)}_{\pm 1}(q,\o^{(0)})$, which are obtained from the
analysis. In particular, slightly below the critical point $\eps_c$, the
unperturbed real frequencies $\O^{(0)}=i\o^{(0)}$ are given by
$|\O^{(0)}| \approx (2/\sqrt{\pi})(T_c-T)$. Thus, in the Vlasov limit,
the homogeneous state with Gaussian-distributed momenta is unstable
below the critical energy, as already noted in Section \ref{sec:HMF}. To obtain the perturbed frequencies $\O$ to lowest order in $\alpha$, we now substitute the unperturbed solutions into Eq. (\ref{lin2}). After a straightforward but lengthy algebra, one obtains at an energy slightly below the critical point the perturbed frequencies to be given by
\be
\O \approx |\O^{(0)}|[1+\alpha A], \mathrm{~with~} A=\fr{2\pi^{3/2}}{\sqrt{3}}\left(1-\fr{1}{\sqrt{5}}\right).
\ee
This equation suggests that to leading order in $\alpha$, the frequencies $\O$ are real for energies just below the critical value and vanish at the critical point. Thus, a homogeneous state with Gaussian-distributed momenta is linearly unstable under the Boltzmann equation at energies just below the critical point and neutrally stable at the critical point.

In the light of the above calculation, we may now analyze the evolution
of magnetization in the generalized HMF model while starting from a
water-bag initial condition. The two timescales that govern the time
evolution of the  magnetization are (i) the scale over which collisions
occur, given by $\alpha^{-1}$, and (ii) the scale $\sim N^\gamma$, over which finite-size effects add corrections to the Boltzmann equation. The interplay between the two timescales may be naturally analyzed by invoking a scaling approach, as we demonstrate below. 

\begin{figure}
\centering
\includegraphics[width=70mm]{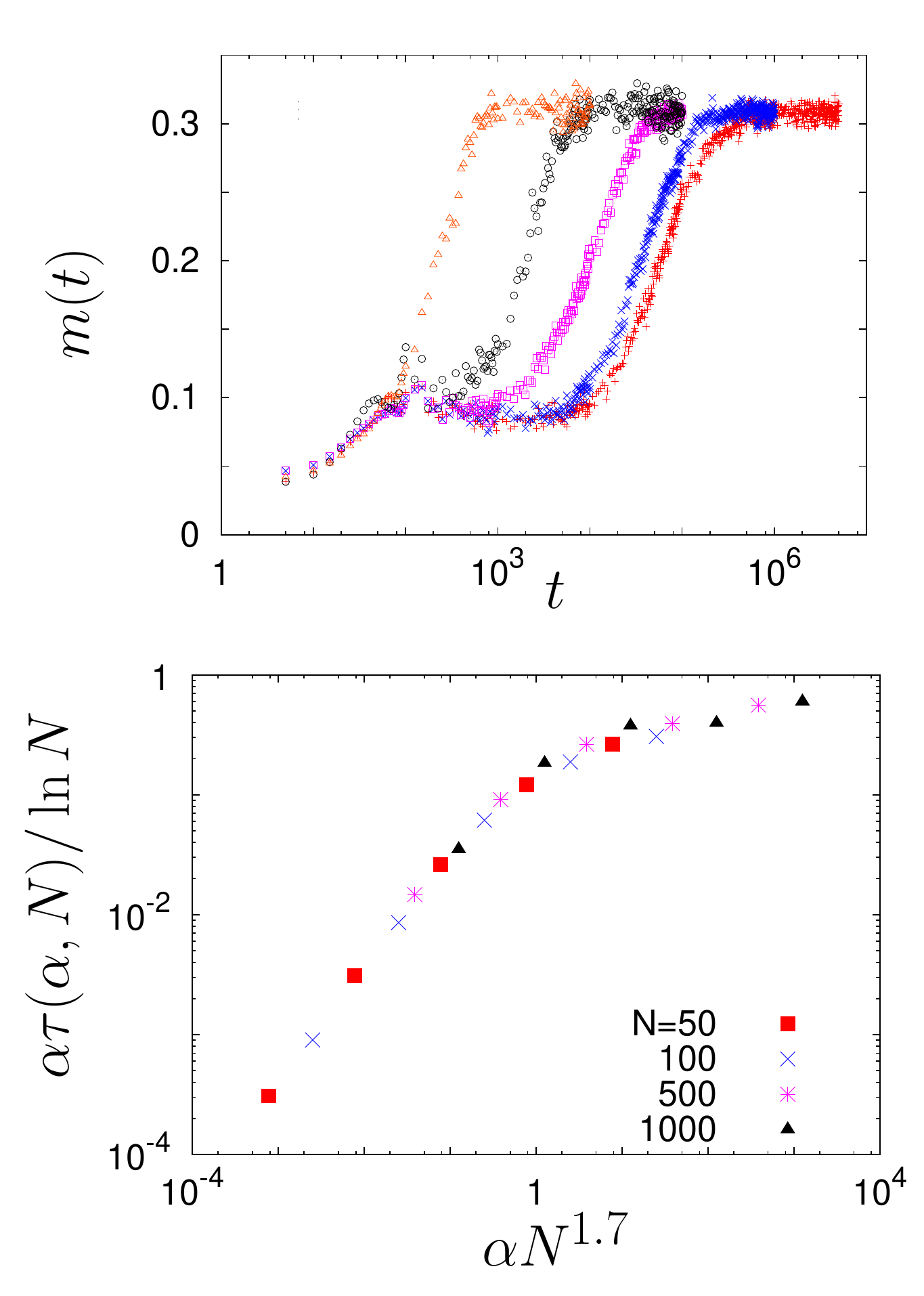}
\caption{(Upper panel) Magnetization versus time for $N=500$ at $\eps=0.69$
and for $\alpha$ values (right to left) $10^{-6}, 10^{-5},
10^{-4},10^{-3}$, and $10^{-2}$. With increasing $\alpha$, one can observe a faster
relaxation towards equilibrium. (Lower panel) $\alpha \tau(\alpha,
N)/\ln N$ versus $\alpha N^\gamma$, showing scaling collapse in
accordance with Eq. (\ref{scaling}). Here, we have $\eps=0.69$.}
\l{fig}
\end{figure}

For $\alpha^{-1} \ll N^\gamma$, and times $\alpha^{-1} \ll t \ll N^\gamma$, the system size is effectively infinite and the evolution follows the Boltzmann equation. Here, frequent collisions at short times drive the momentum distribution towards a Gaussian. As noted above, until this happens, the initial magnetization does not change in time. Over the time the momentum distribution becomes Gaussian, the instability of such a state under the Boltzmann equation leads to a fast relaxation towards equilibrium, similar to the result for the Vlasov-unstable regime. The asymptotic behavior of the magnetization is thus
\be
m(t) \sim \fr{1}{\sqrt{N}}e^{\alpha t};~N^\gamma \gg t \gg \alpha^{-1}.
\l{mag1}
\ee
By requiring that $m(t)$ acquires a value of $O(1)$, the above equation gives the relaxation time $\tau_\mathrm{S}$, determined by the stochastic process, as $\tau_\mathrm{S} \sim \ln N/\alpha$.
In the opposite limit, $\alpha^{-1} \gg N^\gamma$, collisions are
infrequent, and therefore, the process that drives the momentum
distribution to a Gaussian is delayed. The magnetization stays close to
its initial value, and relaxes only over the time $\sim N^\gamma$, over which finite-size effects come into play. Here, similar to the result for the Vlasov-stable regime, the magnetization at late times behaves as
\be
m(t) \sim \fr{1}{\sqrt{N}}e^{t/N^\gamma};~\alpha^{-1} \gg t \gg N^\gamma.
\l{mag2}
\ee
This equation gives the relaxation time $\tau_\mathrm{D}$, determined by
the deterministic process, as $\tau_\mathrm{D} \sim N^\gamma \ln N$.
Interpolating between the above two limits of the timescales, one
expects the relaxation time $\tau(\alpha,N)$ to obey
$\tau^{-1}=\tau_\mathrm{S}^{-1}+\tau_\mathrm{D}^{-1}$, yielding
$\tau(\alpha, N) \sim \ln N/(\alpha + 1/N^\gamma)$. More generally, this suggests a scaling form
\be
\tau(\alpha, N) \sim \fr{\ln N}{\alpha}g(\alpha N^{\gamma}),
\l{scaling}
\ee
where, consistent with Eqs. (\ref{mag1}) and (\ref{mag2}), the scaling function $g(x)$ behaves as follows: $g(x) \sim x$ for $x \ll 1$, while $g(x) \rightarrow$ constant for $x \gg 1$. Equation (\ref{scaling}) implies that for fixed $N$, the relaxation
time of the water-bag initial state exhibits a crossover, from being of
order $N^\gamma \ln N$ (corresponding to QSSs) for $\alpha \ll
1/N^\gamma$ to being of order $\ln N$ for $\alpha \gg 1/N^\gamma$. This
brings us to the main conclusion of this Subsection: In the presence of
collisions, the relaxation at long times does not occur over an
algebraically growing timescale, which implies that under noisy
microcanonical evolution, QSSs occur only as a crossover phenomenon and are lost in the limit of long times. 

The above predictions, in particular, the scaling form in Eq.
(\ref{scaling}), may be verified by performing extensive numerical
simulations of the generalized HMF model. The Hamilton equations, Eq.
(\ref{eq:hameq}), may be integrated by using a symplectic fourth-order integrator. In realizing the stochastic process $(p,p',p'') \rightarrow (q,q',q'')$ while conserving the three-particle energy $E$ and momentum $P$, we note that the updated momenta lie on a circle formed by the intersection of the plane $p+p'+p''=P$ and the spherical surface $p^2+p'^2+p''^2=2E$. The radius of this circle is given by $r=\sqrt{2E-P^2/3}$. The new momenta may thus be parametrized in terms of an angle $\phi$ measured along this circle, as $q=(P/\sqrt{3})+r\sqrt{2/3}\cos \phi, q'=(P/\sqrt{3})-(r/\sqrt{6})\cos \phi-(r/\sqrt{2})\sin \phi, q''=(P/\sqrt{3})-(r/\sqrt{6})\cos \phi+(r/\sqrt{2})\sin \phi$. Stochasticity in updates is achieved through choosing the angle $\phi$ uniformly in $[0,2\pi)$.
Following the foregoing scheme, typical time evolution of the
magnetization in the generalized HMF model for $N=500$ and several values of $\alpha$ at an energy
density $\eps=0.69$ are shown in Fig. \ref{fig}(Upper panel). The relaxation time
$\tau(\alpha,N)$ is taken as the time for the magnetization to reach the
fraction $0.8$ of the final equilibrium value (the result, however, is
not sensitive to this choice). At $\eps=0.69$, where the equilibrium
value of the magnetization is $\simeq 0.3$ and $\gamma \simeq 1.7$, we
plot $\alpha \tau(\alpha, N)/\ln N$ versus $\alpha N^\gamma$ to check the
scaling form in Eq. (\ref{scaling}). Figure \ref{fig}(Lower panel) shows an
excellent scaling collapse over several decades; this is consistent with
our prediction for QSSs as a crossover phenomenon under noisy microcanonical dynamics.

\subsection{HMF model generalized to particles moving on a
sphere}
\l{sec:hmf-sphere}

In order to probe the ubiquity of the quasistationary behavior observed
in the HMF model, various extensions of the model have been introduced and analyzed over the
years. For example, the HMF model was
considered with an additional term in the energy that is due to either 
(i) a global anisotropy in the magnetization along the $x$-axis, or,
(ii) an onsite potential. In either case, QSSs were shown to exist in
specific energy ranges, with a relaxation time scaling algebraically
with the system size. A particularly 
interesting generalization of the HMF model is to that of particles moving on
the surface of a sphere rather than on a circle \cite{Gupta:2013}: Consider a system of $N$ interacting particles moving on the surface of a unit
sphere. The generalized coordinates of the $i$-th particle are the spherical polar angles $\th_i \in [0,\pi]$ and $\ph_i
\in [0,2\pi]$, while the corresponding generalized momenta are 
$p_{\th_i}$ and $p_{\ph_i}$. The Hamiltonian of the system is
given by 
\bea
H=\fr{1}{2}\sum_{i=1}^N \Big(p_{\th_i}^2+\fr{p_{\ph_i}^2}{\sin^2
\th_i}\Big)+\fr{1}{2N}\sum_{i,j=1}^N\Big[1-{\bf S}_i\cdot {\bf S}_j\Big].
\l{eq:sphere-hmf-H}
\eea
Here, ${\bf S}_i$ is the vector pointing from the center to the position of the $i$-th particle on the sphere, and 
has the Cartesian components $(S_{ix},S_{iy},S_{iz})=(\sin
\th_i \cos \ph_i,\sin \th_i \sin \ph_i, \cos \th_i)$. Regarding the vector ${\bf S}_i$ as the classical Heisenberg spin
vector of unit length, the interaction term in Eq.
(\ref{eq:sphere-hmf-H}) has a form similar to that
in a mean-field Heisenberg model of magnetism. However, unlike the latter case, the
Poisson bracket between the components of ${\bf S}_i$'s is
identically 
zero. Relative to the HMF model, the model (\ref{eq:sphere-hmf-H}) is defined on a larger
phase space with each particle characterized by two positional degrees
of freedom rather than one.

The interaction term in Eq. (\ref{eq:sphere-hmf-H}) tries to cluster the particles, and is
in competition with the kinetic energy term (the term involving
$p_{\th_i}$ and $p_{\ph_i}$) that has the opposite
effect.  The degree of clustering is conveniently measured by the
``magnetization" vector ${\bf m}=(m_x,m_y,m_z)\equiv\sum_{i=1}^N {\bf S}_i/N$. In the BG equilibrium state, the system exhibits a
continuous phase transition at the critical energy density
$\eps_c\equiv5/6$, between a low-energy clustered (``magnetized") phase in
which the particles are close together on the sphere, and a high-energy
homogeneous (``non-magnetized") phase in which the particles are uniformly distributed on
the sphere. As a function of the energy,
the magnitude of ${\bf m}$, i.e., $m=\sqrt{m_x^2+m_y^2+m_z^2}$, decreases continuously from unity at zero energy density to zero at $\eps_c$, and
remains zero at higher energies. The mentioned phase transition
properties may be derived by following a procedure similar to the one
invoked in Section \ref{sec:HMF}.

 The time evolution of the system (\ref{eq:sphere-hmf-H}) follows the usual Hamilton
equations of motion derived from the Hamiltonian (\ref{eq:sphere-hmf-H}).
The issue of how the system while starting far from
equilibrium and evolving under the Hamilton equations relaxes to the
equilibrium state may be investigated by studying 
the Vlasov equation for the evolution of the single-particle phase space
density. Let $f(\th,\ph,p_\th,p_\ph,t)$ be the probability density in
this phase space, such that $f(\th,\ph,p_\th,p_\ph,t){\rm d}\th
{\rm d}\ph {\rm d}p_\th {\rm d}p_\ph$ gives the probability at time $t$ to find the
particle with its generalized coordinates in $(\th,\th+{\rm d}\th)$ and
$(\ph,\ph+{\rm d}\ph)$,
and the corresponding momenta in $(p_\th,p_\th+{\rm d}p_\th)$ and
$(p_\ph,p_\ph+{\rm d}p_\ph)$. The Vlasov equation reads \cite{Gupta:2013} 
\bea
&&\fr{\partial f}{\partial t}+p_\th\fr{\partial f}{\partial
\th}+\fr{p_\ph}{\sin^2\th}\fr{\partial f}{\partial \ph}+\Big(\fr{p^2_{\ph}
\cos \th}{\sin ^3\th}+m_x \cos
\th \cos \ph+m_y \cos \th \sin \ph-m_z \sin \th\Big)\fr{\partial f}{\partial
p_\th}\nonumber \\
&&+(-m_x \sin \th \sin \ph+m_y \sin \th \cos \ph)\fr{\partial
f}{\partial p_\ph}=0;
\l{eq:sphere-hmf-Vlasov}\\
&&(m_x,m_y,m_z)=\int {\rm d}\th {\rm d}\ph {\rm d}p_\th {\rm d}p_\ph
~(\sin \th\cos \ph,\sin \th \sin \ph,\cos
\th)f.
\eea
It is easily verified that any distribution
$f^{(0)}(\th,\ph,p_\th,p_\ph)=\Phi(e(\th,\ph,p_\th,p_\ph))$, with arbitrary
function $\Phi$, and $e$ being the single-particle energy,
\be
e(\th,\ph,p_\th,p_\ph)=\fr{1}{2}\Big(p_{\th}^2+\fr{p_{\ph}^2}{\sin^2
\th}\Big)-m_x \sin \th \cos \ph-m_y \sin \th \sin \ph-m_z \cos \th,
\l{singlee}
\ee
is stationary under the Vlasov dynamics (\ref{eq:sphere-hmf-Vlasov}). The magnetization
components $m_x,m_y,m_z$ are determined self-consistently. As a specific
example, consider a stationary state that is non-magnetized, that is,
$m_x=m_y=m_z=0$, and $f^{(0)}(\th,\ph,p_\th,p_\ph)$ is given by
\be
f^{(0)}(\th,\ph,p_\th,p_\ph)=\left\{ 
\begin{array}{ll}
               &\fr{1}{2\pi}\fr{1}{\pi}A \mbox{~~if~}
             \fr{1}{2}\Big(p_{\th}^2+\fr{p_{\ph}^2}{\sin^2
               \th} \Big)<E; \\
               &\th \in [0,\pi],\ph \in [0,2\pi],
               A,E\ge 0, \\
               &0,  \mbox{~~~~~~~~otherwise}.
               \end{array}
        \right. \\
\l{eq:sphere-hmf-f0}        
\ee
The state (\ref{eq:sphere-hmf-f0}) is a straightforward generalization of the water-bag initial condition for the HMF model. The parameters $A$ and $E$ are related through the normalization
condition, $\int_{0}^{\pi}{\rm d}\th\int_{0}^{2\pi}{\rm d}\ph
\int_{\Omega}{\rm d}p_\th {\rm d}p_{\ph}f^{(0)}=1$, where the integration over $p_\th$ and $p_\ph$ is over the domain
$\Omega \equiv \Theta\Big(2E-p_{\th}^2-p_{\ph}^2/\sin^2\th\Big)$,
with $\Theta(x)$ denoting the unit step function. One gets $E=1/(4A)$, while the conserved energy density
$\eps=1/2+\int_{0}^{\pi}{\rm d}\th\int_{0}^{2\pi}{\rm d}\ph
\int_\Omega {\rm d}p_\th {\rm
d}p_{\ph}(1/2)\Big(p_{\th}^2+p_{\ph}^2/\sin^2\th\Big)f^{(0)}$
is related to $E$ as $\eps=(E+1)/2$.

Analyzing the linear stability of the stationary state (\ref{eq:sphere-hmf-f0}) under the
Vlasov dynamics (\ref{eq:sphere-hmf-Vlasov}), it may be shown that for
energies $\eps>\eps^*=2/3$, the non-magnetized state
(\ref{eq:sphere-hmf-f0}) is linearly
stable, and is hence a QSS. In a finite system, the QSS eventually
relaxes to BG equilibrium over a very long timescale, which, considering
the magnetization as an indicator for the relaxation process for
energies $\eps<\eps_c$, grows algebraically with the
system size as $N^\gamma;~\gamma>0$; this is demonstrated by numerical simulation
results in Fig. \ref{fig:sphere-hmf}. For energies $\eps < \eps^*$, the state (\ref{eq:sphere-hmf-f0}) is linearly
unstable, and is thus not a QSS; in this case, numerical simulations
show that the system exhibits a fast relaxation towards BG equilibrium over a timescale that grows with the system size
as $\ln N$. These features of a linearly unstable and a linear stable
regime of a non-magnetized Vlasov-stationary state, with a QSS emerging in the latter
case, remain unaltered on adding a term
to the Hamiltonian (\ref{eq:sphere-hmf-H}) that accounts for a global anisotropy
in the magnetization. We note that a non-mean-field version of the model 
was studied in Ref. \cite{Cirto:2015} in the context of the existence of
QSSs.

\def\figsubcap#1{\par\noindent\centering\footnotesize(#1)}
\begin{figure}[!ht]
\centering
\includegraphics[width=3in]{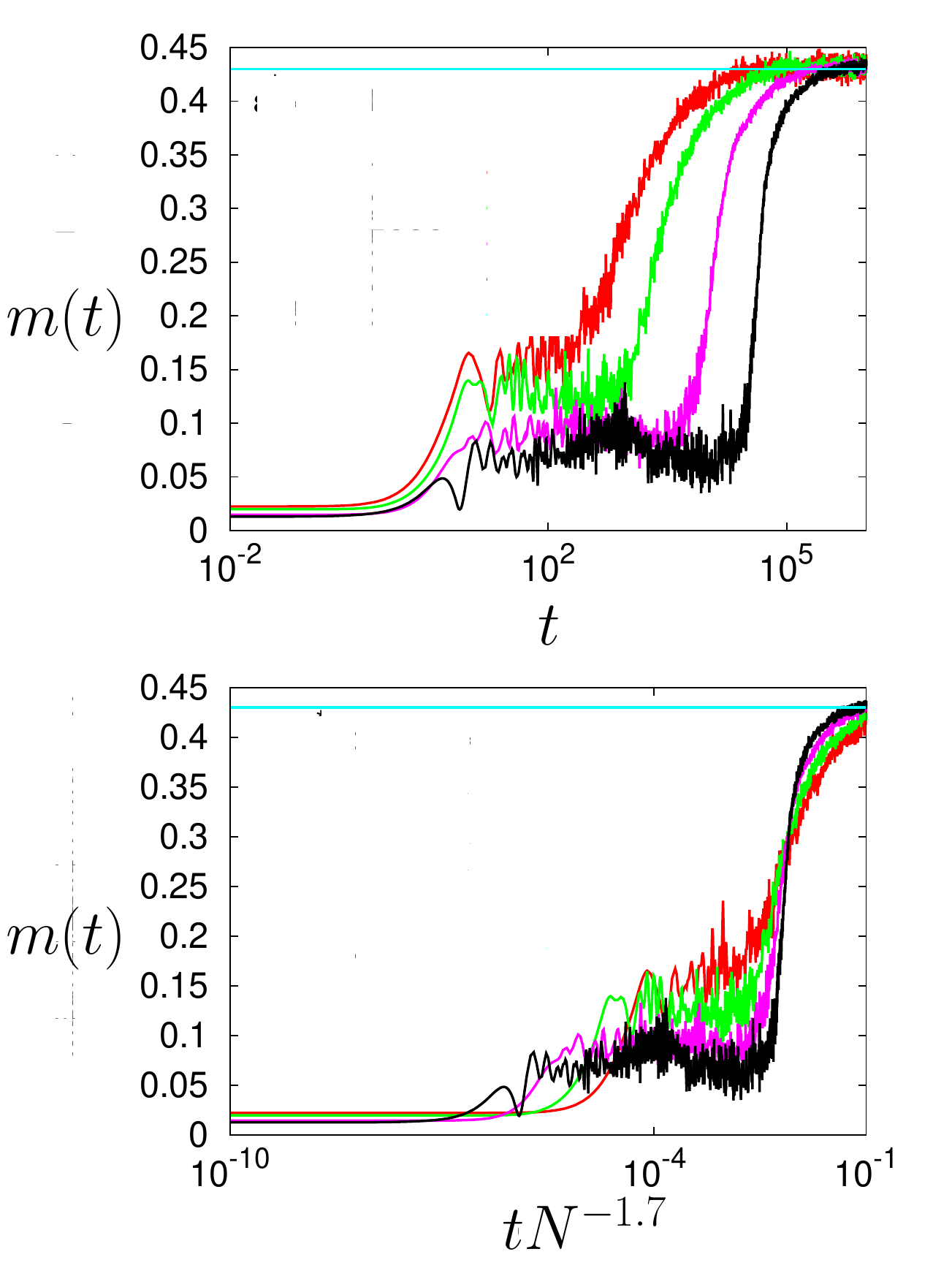}
  \caption{For the model (\ref{eq:sphere-hmf-H}), the figures show
  numerical simulation results for the magnetization $m(t)$ as a function of time
(Upper panel), and as a function of time scaled by $N^{1.7}$ (Lower
panel) in the
Vlasov-stable phase ($\eps> \eps^*$). The energy density is $\eps=0.7$.
The blue line indicates the BG equilibrium value. The
figures suggest the existence of a QSS with a lifetime scaling with the
system size as $N^{1.7}$.}
  \l{fig:sphere-hmf}
\end{figure}

\subsection{A long-range model of spins}
\l{sec:spin-model}

The ubiquity of QSSs may be tested in a dynamics very different from
the particle dynamics of either the HMF model or any of its
generalizations, including the model (\ref{eq:sphere-hmf-H}), namely,
within classical spin dynamics of an anisotropic Heisenberg model with
mean-field interactions \cite{Gupta:2011,Barre:2014}. The model
comprises $N$ globally coupled three-component Heisenberg spins of unit length, ${\mathbf S}_i=(S_{ix},S_{iy},S_{iz})$,
$i=1,2,\ldots,N$. In terms of spherical polar angles $\th_i \in
[0,\pi]$ and $\phi_i \in [0,2\pi]$ for the orientation of the $i$-th
spin, one has $S_{ix}=\sin \th_i \cos \ph_i, S_{iy}=\sin \th_i \sin
\ph_i, S_{iz}=\cos \th_i$. The Hamiltonian of the model is given by
\be
H=-\fr{J}{2N}\sum_{i,j=1}^N {\bf S}_i \cdot {\bf
S}_j+D\sum_{i=1}^N S^2_{iz},
\l{eq:spin-H}
\ee
where the first term with $J
> 0$ describes a ferromagnetic mean-field like coupling, while the
last term gives the energy due to a local anisotropy. We take $D>0$
such that at equilibrium, the energy is lowered by having the
magnetization ${\bf m}\equiv(1/N)\sum_{i=1}^N{\bf S}_i$ pointing in the $xy$
plane. The model (\ref{eq:spin-H}) has an equilibrium phase diagram with a continuous
transition from a low-energy magnetic phase ($m \ne 0$) to a high-energy
non-magnetic phase ($m=0$) across the critical energy density
$\eps_c=D\left(1-2/\beta_c\right)$, where $\beta_c$ satisfies
$2/\beta_c=1-1/(2\beta_cD)+\exp(-\beta_c D)/(\sqrt{\pi\beta_cD}
\mathrm{Erf}[\sqrt{\beta_cD}])$, with ${\rm Erf}(x)$ being the error
function. The derivation of these properties is detailed in Ref. \cite{Gupta:2011}. 

The time evolution of the model (\ref{eq:spin-H}) is governed
by the set of equations
\be
\fr{{\rm d}{\bf S}_i}{{\rm d}t}=\{{\bf S}_i,H\}; ~~~~i=1,2,\ldots,N.
\l{eq:spin-eom}
\ee
Here the Poisson bracket $\{A,B\}$ for two functions of the spins
is obtained by noting that suitable canonical variables for a
classical spin are $\phi$ and $S_z$, so that in our model,
$\{A,B\}\equiv \sum_{i=1}^N (\partial A/\partial \phi_i
\partial B/\partial S_{iz}-\partial A/\partial S_{iz}
\partial B/\partial \phi_i)=\sum_{i=1}^N {\bf S}_i \cdot \partial
A/\partial {\bf S}_i\times \partial B/\partial {\bf S}_i$,
using which one obtains straightforwardly
\bea
&&\frac{{\rm d}S_{ix}}{{\rm d}t}=S_{iy}m_z-S_{iz}m_y-2DS_{iy}S_{iz}, \l{eq:spin-eqnmotionx} \\
&&\frac{{\rm d}S_{iy}}{{\rm d}t}=S_{iz}m_x-S_{ix}m_z+2DS_{ix}S_{iz}, \l{eq:spin-eqnmotiony}\\
&&\frac{{\rm d}S_{iz}}{{\rm d}t}=S_{ix}m_y-S_{iy}m_x. \l{eq:spin-eqnmotionz}
\eea
From Eq. (\ref{eq:spin-eqnmotionz}), one finds by summing over $i$ that
$m_z$ is a constant of motion. The motion also conserves the total
energy and the length of each spin. 

To study the relaxation to equilibrium while starting far from it, one
analyzes as usual the Vlasov equation for the evolution of the single-spin phase space
density. Denoting the latter by $f(\th,\ph,t)$, with $f(\th,\ph,t)\sin
\th {\rm d}\th {\rm d}\ph$ giving the probability to find a spin with
its angles between $\th$ and $\th+{\rm d}\th$ 
and between $\ph$ and $\ph+ {\rm d}\ph$ at time $t$, the Vlasov equation may
be shown to be of the form \cite{Gupta:2011}
\be
\fr{\partial f}{\partial t}=\Big[m_y\cos \ph-m_x\sin \ph\Big]\fr{\partial f}{\partial \th}-\Big[m_x\cot \th\cos \ph+m_y\cot \th\sin \ph
-m_z+2D\cos \th\Big]\fr{\partial f}{\partial\phi}. 
\l{eq:spin-Vlasov}
\ee
In the above equation, the magnetization components are given by $(m_x,m_y,m_z)=\int\sin\th'{\rm d}\th'{\rm d}\ph'
(\sin \th'\cos \ph',\sin \th' \sin \ph',\cos
\th')f(\th',\ph',t)$.

Consider an initial state prepared by sampling independently
for each of the $N$ spins the angle $\phi$ uniformly over $[0,2\pi]$
and the angle $\th$ uniformly over an arbitrary interval symmetric about
$\pi/2$. Such a 
state will have the distribution
\be
f(\th,\ph,0)=\fr{1}{2\pi}p(\th),
\l{eq:spin-waterbag}
\ee
with $p(\th)$, the distribution for $\th$, given by
\be
p(\th)=\left\{
\begin{array}{ll}
               \fr{1}{2\sin a} & \mbox{if $\th \in \left[\fr{\pi}{2}-a,\fr{\pi}{2}+a\right]$}, \\
               & \\
               0 & \mbox{otherwise}.
               \end{array}
        \right. \\
\l{eq:spin-pth}
\ee
Here, $a>0$ is a given parameter. The state (\ref{eq:spin-waterbag}) is analogous to the water-bag state studied in the context of the HMF model.
It is easily verified that this non-magnetic state has the energy
$\eps=(D/3)\sin^2 a$, and that the state is stationary under the
Vlasov dynamics (\ref{eq:spin-Vlasov}).

A linear stability analysis of the state (\ref{eq:spin-waterbag}) under
the Vlasov dynamics (\ref{eq:spin-Vlasov}) shows that the state is
linearly stable for energies $\eps>\eps^*\equiv D/(3+12 D)$, and is thus a
QSS. In this case, in a finite system, such a state eventually relaxes to
BG equilibrium; studying the time evolution of the magnetization to
monitor this relaxation for energies $\eps<\eps_c$, it may be seen that
the relaxation occurs on a timescale $\sim N^{\gamma}$, with $\gamma >0$, see Fig. \ref{fig:spin}. A detailed analytical study of the Lenard-Balescu operator that accounts
at leading order for the finite-size effects driving the relaxation of
the QSSs was taken up in Ref. \cite{Barre:2014}, and it was demonstrated that indeed
corrections at leading order are identically zero, so that relaxation
has to occur over a time longer than of order $N$, in agreement with the numerical results.
For $\eps <
\eps^*$, when the water-bag state is linearly unstable, the magnetization
shows a relaxation from the initial value over a timescale $\tau(N) \sim
\ln N$, see Ref. \cite{Gupta:2011}.

\begin{figure}
\centering
\includegraphics[width=3.0in]{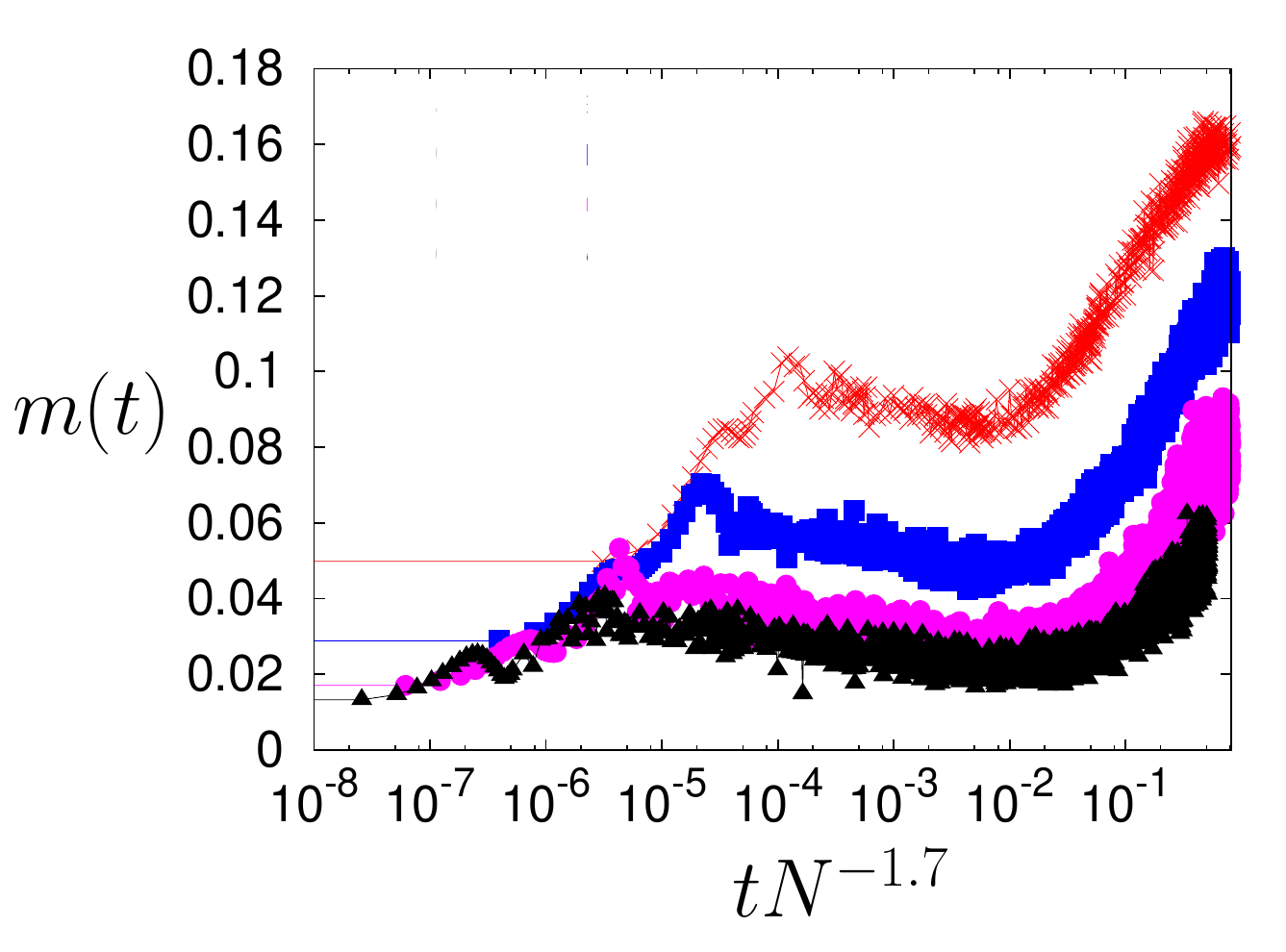}
\caption{For the model (\ref{eq:spin-H}), the figure shows numerical
simulation results for the magnetization $m(t)$ as
a function of $tN^{-1.7}$ with energy density
$\eps=0.24>\eps^*$, the parameter $D=15$, and for systems of size $N=300,1000,3000, 5000$ (top to
bottom).
The figure suggests a QSS
life-time $\tau(N) \sim N^{1.7}$.}
\l{fig:spin}
\end{figure}

\section{Driving a long-range system out of thermal equilibrium:
Temperature inversion and cooling}
\l{sec:temperature-inversion}

What happens when an isolated macroscopic long-range system in thermal equilibrium is
momentarily disturbed, e.g., by an impulsive force or a ``kick''?
How different from an equilibrium state
is the stationary state the system relaxes to after the kick? Are there ways to
characterize it, e.g., by unveiling some of its general features?
These questions were addressed in detail in a recent series of papers
\cite{Casetti:2014,Teles:2015,Teles:2016,Gupta:2016}, demonstrating that when the equilibrium state is spatially
inhomogeneous, the system after the kick relaxes to a QSS that is
characterized by a non-uniform temperature profile in space.
In short-range systems, by contrast, a
non-uniform temperature profile may only occur when the system is
actively maintained out of equilibrium, e.g., by a boundary-imposed
temperature gradient, to counteract collisional effects. In addition to
a non-uniform temperature profile, in a long-range system, the QSS
attained following the kick generically exhibits a remarkable phenomenon
of temperature inversion. Namely, the temperature and density profiles
as a function of space are anticorrelated, that is, denser parts of the
system are colder than dilute ones. Temperature inversion is observed in
nature, e.g., in interstellar molecular clouds and especially in the
solar corona, where temperatures around $10^6$ K that are three orders
of magnitude larger than the temperature of the photosphere are
attained.

\begin{figure}[!ht]
\begin{center}
\includegraphics[width=120mm]{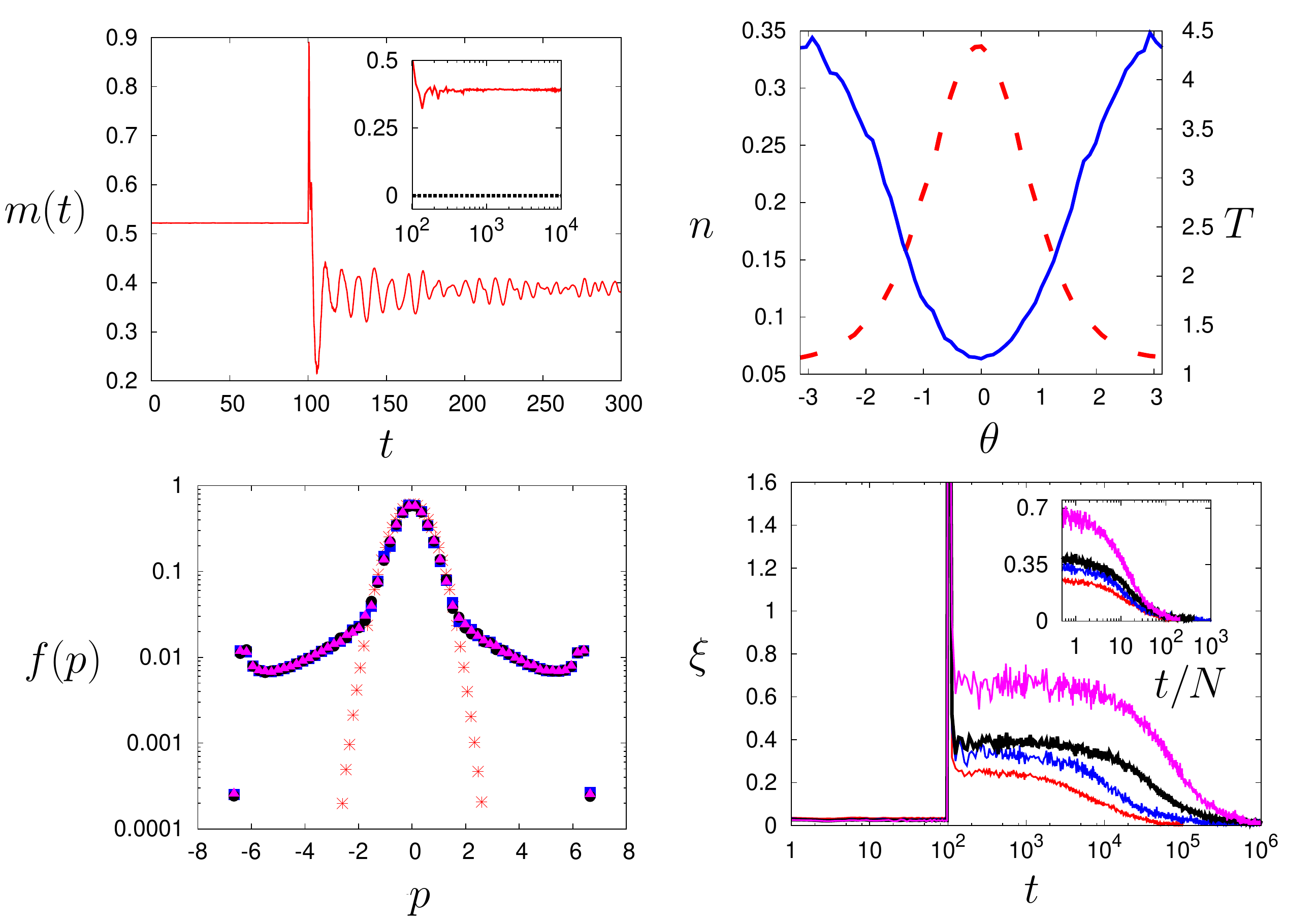}
  \caption{In the HMF model, while starting with thermal equilibrium at
  temperature $T = 0.4244$ with corresponding equilibrium magnetization $m_x =0.521$
  and $m_y = 0$, the system is let evolve until $t=t_0=100$, and then kicked out of equilibrium
by applying an external magnetic field $h=10.0$ along the $x$ direction
for times $t_0 < t < t_0 + 1$. The first three panels show molecular dynamics
simulation results for $N=10^7$ for (First panel) Time evolution of the magnetization $m$
(solid red line); here the inset shows for longer times $m(t)$ compared to the equilibrium value $m_{\text{eq}} = 0$ (dotted black
  line) at the same energy; (Second panel) Temperature profile $T(\theta)$ (blue
  solid line) and density profile $n(\theta)$ (red dashed line)
  measured in the QSS obtained at $t = 10^4$; (Third panel) Momentum distribution
  $f(p)$ at $t=0$ (red crosses), $t = 5\times10^2$ (blue squares), $t =
  10^3$ (black circles), $t = 10^4$ (purple triangles); that the
  distributions for the last three cases are indistinguishable implies
  that the system is in a stationary state (and is in fact in a QSS, see
  discussions in the main text). The fourth panel shows
  time evolution of the space-integrated distance
  $\xi$ of the instantaneous temperature from the equilibrium
  temperature at the same energy, for different values of $N$
  increasing from bottom to top as $N = 5\times10^2$ (red), $N = 10^3$
  (blue), $N = 2.5\times10^3$ (black), $N = 5\times 10^3$ (purple). The inset of the fourth panel shows $\xi$ as a function of $t/N$. }
  \l{fig:temp-inversion}
\end{center}
\end{figure}

To demonstrate the claim of temperature inversion, the dynamical evolution of the HMF system kicked out of thermal
equilibrium may be studied via molecular dynamics (MD) simulations
involving numerical integration of its equations of motion. As an
illustrative example, the system is initially prepared in thermal equilibrium at temperature $T = 0.4244$ with corresponding equilibrium magnetization
  $m_x =m_0=0.521$ and $m_y = 0$, let evolve until $t = t_0>0$, and then kicked out of equilibrium
by applying during a short time $\tau$ an external magnetic field $h$ along the $x$ direction;
thus, for $t_0 < t < t_0 + \tau$, the Hamiltonian (\ref{eq:hmf-H}) is
augmented by the term $H_h = - h\sum_{i=1}^N \cos \theta_i$.
Here, we present results for $t_0 = 100$, $\tau = 1$, $h = 10$, and
$N=10^7$. After the kick, the magnetization starts oscillating, but
eventually damps down to a stationary value smaller than $ m_0$. A typical time
evolution of the magnetization is shown in Fig.
\ref{fig:temp-inversion}, First panel. The stationary state reached after the damping of the oscillations is a QSS.
The nonequilibrium character of this state is shown by the fact that the temperature profile
\be
T(\theta) \equiv \frac{\int_{-\infty}^\infty dp\, p^2 f(\theta,p)}{\int_{-\infty}^{\infty} dp\, f(\theta,p)}
\l{tempprofile}
\ee
is non-uniform, and there is temperature inversion, as shown in Fig.
\ref{fig:temp-inversion}, Second panel, where $T(\theta)$ is plotted together with the density profile 
\be
n(\theta) \equiv \int_{-\infty}^{\infty} dp\, f(\theta,p).
\l{densprofile}
\ee
Here, $f(\theta,p)$ is the usual single-particle phase space density. The temperature profile indeed remains essentially the same for the
whole lifetime of the QSS, as may be checked by measuring an integrated distance $\xi$ between the
actual temperature profile and the constant equilibrium one,
$T_{\text{eq}}$, at the same energy, as follows:
\be
\xi(t) \equiv \int_{-\pi}^\pi \left| T(\theta,t) - T_{\text{eq}} \right| d\theta.
\l{eq:xi}
\ee
In Fig. \ref{fig:temp-inversion}, Third panel, we show that the momentum
distribution in the QSS reached after the kick develops supra-thermal
tails, while in the fourth panel, $\xi(t)$ is plotted for systems with different
values of $N$ kicked with the same $h = 10$ at $t_0 = 100$ for a duration $\tau = 1$. After the kick, $\xi(t)$ oscillates and then
reaches a plateau whose duration grows with $N$, as expected for a QSS.
The inset of Fig. \ref{fig:temp-inversion}, Fourth panel, shows that if times
are scaled by $N$, the curves reach zero at the same time, consistently
with the lifetime of an inhomogeneous QSS being proportional to $N$.

\section{Conclusions}
\l{sec:conclusions}

In this brief contribution, we offered an overview of properties of long-range interacting (LRI) systems. We exclusively
focussed on systems for which the long-time stationary state is in
equilibrium. Because of lack of space, we could not cover the even
richer static and dynamics properties exhibited by systems that have a non-equilibrium
stationary state \cite{p1,p2,p3,p4,p5,p6,p7,p8,p9,p10}. LRI systems present a particularly exciting
area of research due to the possibility to develop theoretical tools
that effectively combine and adapt methods and techniques from diverse fields, but also in the wake of new experimental realizations of LRI
systems that offer the possibility to directly test the predictions obtained in
theory. We hope that this contribution will serve as an invitation
to young (and old) minds to delve into the exciting world of long-range interactions.

\section*{Acknowledgments}
We would like to thank all our collaborators for having fruitful discussions
and enjoyable collaborations over the years on topics covered in this
article: Julien Barr\'{e}, Fernanda P. C. Benetti, Freddy Bouchet, Alessandro Campa, Lapo
Casetti, Pierre-Henri Chavanis, Pierfrancesco di Cintio,
Thierry Dauxois, Maxim Komarov, Yan Levin, David Mukamel, Cesare
Nardini, Renato Pakter, Aurelio
Patelli, Arkady Pikovsky, Max
Potters, Tarcisio N. Teles and Yoshiyuki Y. Yamaguchi.



\begin{thebibliography}{9}

\bibitem{Dauxois:2002}{\it Dynamics and Thermodynamics of Systems with
Long-Range Interactions, Lecture Notes in Physics, vol. 602}, edited by
T. Dauxois, S. Ruffo, E. Arimondo, and M. Wilkens (Springer,
Berlin, 2002).

\bibitem{Campa:2008}{\it Dynamics and Thermodynamics of Systems with
Long-range Interactions: Theory and Experiment}, edited by A. Campa, A. Giansanti,
G. Morigi, and F. Sylos Labini, AIP Conference Proceedings 970 (2008).

\bibitem{Ruffo:2008}S. Ruffo, Eur. Phys. J. B {\bf 64}, 355 (2008).

\bibitem{Campa:2009}A. Campa, T. Dauxois, and S. Ruffo, Phys. Rep. {\bf 480}, 57 (2009).

\bibitem{Dauxois:2009}{\it Long-range interacting systems},
edited by T. Dauxois, S. Ruffo, and L. Cugliandolo (Oxford University Press, Oxford, 2009).

\bibitem{Bouchet:2010}F. Bouchet, S. Gupta, and D. Mukamel, Physica A {\bf 389}, 4389 (2010).

\bibitem{Dauxois:2010}Journal of Statistical Mechanics: Theory and Experiment Topical issue:
{\it Long-Range Interacting Systems}, edited by T. Dauxois and S. Ruffo (2010).

\bibitem{Campa:2014}A. Campa, T. Dauxois, D. Fanelli,
and S. Ruffo, {\it Physics of Long-Range Interacting Systems}, (Oxford University Press, Oxford, 2014)

\bibitem{Kiessling:1997}M. Kiessling and J. L. Lebowitz, Letters in
Mathematical Physics {\bf 42}, 43 (1997).

\bibitem{Barre:2001}J. Barr\'{e}, D. Mukamel, and S. Ruffo, Phys. Rev. Lett.
{\bf 87}, 030601 (2001).

\bibitem{Ellis:2002}R. S. Ellis, K. Haven, and B. Turkington,
Nonlinearity {\bf 15}, 239 (2002).

\bibitem{Pikovsky:2014}A. Pikovsky, S. Gupta, T. N. Teles, F. P. C.
Benetti, R. Pakter, Y. Levin, and S. Ruffo, Phys. Rev.
E {\bf 90}, 062141 (2014).

\bibitem{Dubin:2010}D. H. E. Dubin, in {\it Long-Range Interacting
Systems}, edited by T. Dauxois, S. Ruffo, and L. F. Cugliandolo (Oxford
University Press, Oxford, 2010).

\bibitem{Bramwell:2010}S. T. Bramwell, in {\it Long-Range Interacting
Systems}, edited by T. Dauxois, S. Ruffo, and L. F. Cugliandolo (Oxford
University Press, Oxford, 2010).

\bibitem{Barre:2004}J. Barr\'{e}, T. Dauxois, G. De Ninno, D. Fanelli,
and S. Ruffo, Phys. Rev. E {\bf 69}, 045501 (R) (2004).

\bibitem{Chavanis:2008}P.-H. Chavanis, in {\it Dynamics and Thermodynamics of Systems with
Long-range Interactions: Theory and Experiment}, edited by A. Campa, A. Giansanti,
G. Morigi, and F. Sylos Labini, AIP Conference Proceedings 970 (2008).

\bibitem{Mukamel:2005}D. Mukamel, S. Ruffo, and N. Schreiber,
Phys. Rev. Lett. {\bf 95}, 240604 (2005).

\bibitem{Antoni:1995}M. Antoni and S. Ruffo, Phys. Rev. E {\bf 52}, 2361
(1995).

\bibitem{Yamaguchi:2004}Y. Y. Yamaguchi, J. Barr\'{e}, F. Bouchet, T.
Dauxois, and S. Ruffo, Physica A {\bf 337}, 36 (2004).

\bibitem{Levin:2014}Y. Levin, R. Pakter, F. B. Rizzato, T. N. Teles, and F.
P. da C. Benetti, Phys. Rep. {\bf 535} 1 (2014).

\bibitem{Anteneodo:1998}C. Anteneodo and C. Tsallis, Phys. Rev. Lett.
{\bf 80}, 5313 (1998).

\bibitem{Gupta:2012}S. Gupta, A. Campa, and Stefano Ruffo, Phys. Rev. E {\bf 86}, 061130
(2012).

\bibitem{Schutz:2014}S. Sch\"{u}tz and G. Morigi, Phys. Rev. Lett. {\bf
113}, 203002 (2014).

\bibitem{Schutz:2015}S. Sch\"{u}tz, S. B. J\"{a}ger, and G. Morigi, Phys. Rev. Lett. {\bf 117}, 083001 (2016).

\bibitem{Jager:2016}S. B. J\"{a}ger, S. Sch\"{u}tz, and G. Morigi, Phys.
Rev. A {\bf 94}, 023807 (2016). 

\bibitem{Gupta:2010-1}S. Gupta and D. Mukamel, Phys. Rev. Lett. {\bf 105}, 040602 (2010).

\bibitem{Gupta:2010-2}S. Gupta and D. Mukamel, J. Stat. Mech.: Theory Exp. P08026 (2010).

\bibitem{Gupta:2013}S. Gupta and D. Mukamel, Phys. Rev. E {\bf 88}, 052137 (2013).

\bibitem{Cirto:2015}L. J. L. Cirto, L. S. Lima, and F. D Nobre, J. Stat.
Mech.: Theory Exp. P04012 (2015).

\bibitem{Gupta:2011}S. Gupta and D. Mukamel, J. Stat. Mech.: Theory Exp. P03015 (2011).

\bibitem{Barre:2014}J. Barr\'{e} and S. Gupta, J. Stat. Mech.: Theory
Exp. P02017 (2014).

\bibitem{Casetti:2014}L. Casetti and S. Gupta, Eur. Phys. J. B {\bf 87}, 91 (2014).

\bibitem{Teles:2015}T. N. Teles, S. Gupta, P. D. Cintio, and L. Casetti, Phys. Rev. E {\bf 92}, 020101(R) (2015).

\bibitem{Teles:2016}
T. N. Teles, S. Gupta, P. D. Cintio, and L. Casetti, Phys. Rev. E {\bf 93}, 066102 (2016). 

\bibitem{Gupta:2016}S. Gupta and L. Casetti, New J. Phys. {\bf 18}, 103051 (2016).

\bibitem{p1}
C. Nardini, S. Gupta, S. Ruffo, T. Dauxois, and F. Bouchet, J. Stat.
Mech.: Theory Exp. L01002 (2012).

\bibitem{p2} S. Gupta, M. Potters, and S. Ruffo,
Phys. Rev. E {\bf 85}, 066201 (2012).

\bibitem{p3}C. Nardini, S. Gupta, S. Ruffo, T.
Dauxois, and F. Bouchet, J. Stat. Mech.: Theory Exp.
P12010 (2012).

\bibitem{p4}S. Gupta, A. Campa, and S. Ruffo, Phys. Rev. E {\bf 89},
022123 (2014).

\bibitem{p5}S. Gupta, T. Dauxois, and S. Ruffo, J. Stat. Mech.:
Theory Exp. P11003 (2013).

\bibitem{p6}M. Komarov, S. Gupta, and A. Pikovsky, EPL
{\bf 106}, 40003 (2014).

\bibitem{p7}S. Gupta, A. Campa, and
S. Ruffo, J. Stat. Mech.: Theory Exp. R08001 (2014).

\bibitem{p8}
A. Campa, S. Gupta, and S. Ruffo, J. Stat. Mech.: Theory Exp. P05011
(2015).

\bibitem{p9}S. Gupta, T. Dauxois, and S. Ruffo, EPL {\bf 113}, 60008
(2016).

\bibitem{p10}
A. Campa and S. Gupta, EPL {\bf 116}, 30003
(2016).


\end{thebibliography}
\end{document}